\documentclass[letterpaper,11pt]{article}

\RequirePackage{amsmath, amsthm, amssymb, mathtools}
\RequirePackage{comment}
\RequirePackage{graphicx}
\RequirePackage{fancyvrb}
\RequirePackage[usenames,dvipsnames,svgnames,table]{xcolor}
    \definecolor{linkcolor}{RGB}{0,0,128}
\RequirePackage[pdfpagelabels,plainpages=false,hypertexnames=true,colorlinks=true,linkcolor=linkcolor,anchorcolor=linkcolor,citecolor=linkcolor,filecolor=linkcolor,menucolor=linkcolor,runcolor=linkcolor,urlcolor=linkcolor,pdfborder={0 0 0}]{hyperref}
\RequirePackage[scale=0.95]{sourcecodepro}
\RequirePackage{mathpazo}

\RequirePackage{dsfont}
\RequirePackage{parskip}
\RequirePackage{tikz}
    \usetikzlibrary{positioning}
\RequirePackage[backend=biber,backref=true,style=numeric,giveninits=true,natbib=true,maxbibnames=12]{biblatex}
\RequirePackage{microtype}
\usepackage{newfloat}
\DeclareFloatingEnvironment[
    fileext=los,
    listname={List of Algorithms},
    name=Algorithm,
    placement=tbhp,
    within={},
]{algorithm}

\pdfmapfile{+sourcecodepro.map}

\theoremstyle{plain}% default
\newtheorem{prototheorem}{Theorem}
\newtheorem{theorem}[prototheorem]{Theorem}
\newtheorem{lemma}[prototheorem]{Lemma}

\newtheorem{corollary}[prototheorem]{Corollary}

\theoremstyle{remark}

\newtheorem{remark}{Remark}

\newtheorem{example}{Example}

\newcommand{\R}{\mathbb{R}}
\DeclareMathOperator{\sign}{sgn}
\newcommand{\prb}{\mathbb{P}}

\title{GIST: Gibbs self-tuning for locally  \\ 
adaptive Hamiltonian Monte Carlo}
\author{Nawaf Bou-Rabee\thanks{Department of Mathematical Sciences, Rutgers University, \href{mailto:nawaf.bourabee@rutgers.edu}{\texttt{nawaf.bourabee@rutgers.edu}}}
\and
Bob Carpenter\thanks{Center for Computational Mathematics, Flatiron Institute, \href{mailto:bcarpenter@flatironinstitute.org}{\texttt{bcarpenter@flatironinstitute.org}}}
\and
Milo Marsden\thanks{Department of Mathematics, Stanford University, 
\href{mailto:mmarsden@stanford.edu}{\texttt{mmarsden@stanford.edu}}}
}

\begin{document}

\maketitle
\begin{abstract}
\noindent
We introduce a novel and flexible framework for constructing locally adaptive Hamiltonian Monte Carlo (HMC) samplers by Gibbs sampling the algorithm's tuning parameters conditionally based on the position and momentum at each step. For adaptively sampling path lengths, our Gibbs self-tuning (GIST) approach encompasses randomized HMC, multinomial HMC, the No-U-Turn Sampler (NUTS), and the Apogee-to-Apogee Path Sampler as special cases.  We exemplify the GIST framework with a novel  alternative to NUTS for locally adapting path lengths, evaluated with an exact Hamiltonian for a high-dimensional, ill-conditioned Gaussian measure and with the leapfrog integrator for a  suite of diverse models.
\end{abstract}

\section{Introduction}

Tuning the parameters of Markov chain Monte Carlo (MCMC) algorithms is critical to their  performance, but remains notoriously difficult in practice.  This challenge is particularly acute for Hamiltonian Monte Carlo (HMC), where the tuning of path length (leapfrog integration time) \cite{HoGe2014,BoSa2017,betancourt2017conceptual,kleppe2022connecting}, step size (time discretization parameter) \cite{BePiRoSaSt2013,betancourt2014optimizing,biron2024automala}, and pre-conditioner (mass matrix) \cite{GiCa2011,kleppe2016adaptive,whalley2024randomized} presents complex trade-offs in computational cost and mixing.   The successful self-tuning of path length provided by the No-U-Turn Sampler (NUTS) has led to its widespread adoption as the default sampler in probabilistic programming languages \cite{carpenter2016stan,salvatier2016probabilistic,nimble-article:2017,ge2018t,phan2019composable}.

In NUTS, the path length is adaptively sampled according to a U-turn avoiding condition, which, roughly speaking, stops the underlying leapfrog trajectory whenever it doubles back \cite{HoGe2014,betancourt2017conceptual}. 
More precisely, the path length in NUTS is generated by a stochastic recursive algorithm that doubles the leapfrog path  at each step in a random direction (forward or backward in time) until a U-turn (in position) is detected. The next state is then randomly selected from among the generated leapfrog iterates, favoring those iterates that are further away from the starting point and those that have lower energy.  This approach ensures reversibility while requiring only linearly many operations and logarithmic memory in the number of leapfrog steps taken.

The powerful benefits of path-length tuning provided by NUTS inevitably raise the question: 
\begin{center} \emph{Can the HMC sampler's other tuning parameters  be similarly self-tuned?} \end{center} This paper answers this question by introducing Gibbs self-tuning (GIST) for HMC: a general framework for adaptively sampling HMC tuning parameters such as  path length, step size, pre-conditioner, etc.\footnote{As discussed in Section~\ref{sec:av}, the approach presented here could also be applied to Metropolis samplers beyond  HMC.}  The resulting class of samplers, called GIST samplers, generalizes and unifies many existing locally adaptive HMC methods, including NUTS.

The core idea behind GIST is to expand the state space by treating tuning parameters as auxiliary variables.  In this enlarged space, a joint distribution is defined by specifying a conditional distribution of the tuning parameters given the current position and momentum.  The GIST sampler alternates between Gibbs updates of momentum and tuning parameters, and a Metropolis step based on a measure-preserving involution.  

A recent application of this framework introduces step-size adaptivity into NUTS, addressing a long-standing open problem:  how to adapt the integrator step size without violating reversibility \cite{BouRabeeCarpenterKleppeMarsden2024,BouRabeeCarpenterKleppeLiu2025}. The first solution, proposed in \cite{BouRabeeCarpenterKleppeMarsden2024}, achieves this through orbit-level adaptivity, selecting a step size once per orbit based on the energy error along the entire orbit.  A follow-up work \cite{BouRabeeCarpenterKleppeLiu2025} advances this further by introducing within-orbit adaptivity, allowing the step size to vary dynamically along the trajectory in response to local energy errors.

The GIST framework draws inspiration from and contributes to two major threads of research in MCMC: the theory of auxiliary variable methods and the theory of Metropolization via involutions. The idea of augmenting the state space with auxiliary variables has a long and fruitful history in MCMC. Pioneering works by Edwards and Sokal \cite{edwards1988generalization}, and Besag and Green \cite{besag1993spatial}, laid the foundation for cluster-based and data-augmentation samplers. Subsequent developments have leveraged this framework to improve mixing and enable global moves across a variety of settings, as surveyed by Higdon \cite{higdon1998auxiliary}, advanced through geometric approaches  by Diaconis and Andersen \cite{Diaconis2007HitandRun}, and recently analyzed through operator-theoretic comparisons by Rudolf and Ullrich \cite{RudolfUllrich2018}. Auxiliary variables now underpin key MCMC algorithms, including HMC itself \cite{DuKePeRo1987,Ne2011}, slice sampling \cite{Neal2003Slice,rudolf2018comparison}, the proximal sampler \cite{pmlr-v134-lee21a,pmlr-v178-chen22c}, and non-reversible generalizations of HMC \cite{neal2020non}. 

GIST also builds on a recent conceptually elegant framework for constructing Metropolis-adjusted kernels using deterministic involutions.  A foundational contribution in this direction is the work of Andrieu, Lee, and Livingstone \cite{AndrieuLeeLivingstone}.  Let $(\mathbb{A},\mathcal{A},\hat{\mu})$ be a probability space equipped with an involution $G : \mathbb{A} \to \mathbb{A}$.  For any acceptance function $a : [0, \infty) \to [0,1]$ satisfying the symmetry condition $a(r) = r a(1/r)$ for all $r>0$, they prove that the kernel
\[
\hat{\pi}(x, dy) = \alpha(x) \delta_{G(x)}(dy) + (1 - \alpha(x)) \delta_{x}(dy) \;, \quad \alpha(x) = a \left( \frac{d \hat{\mu} \circ G}{d\hat{\mu}}(x) \right) \;,
\]
is reversible with respect to $\hat{\mu}$.   As special cases, their framework recovers both the Metropolis rule ($a(r) = 1 \wedge r$) and the Barker rule ($a(r) = 1/(1+r)$).  As in our setting, they study auxiliary variable constructions, where the state space is a product $\mathbb{A} = \mathbb{S} \times \mathbb{V}$ of the original space $\mathbb{S}$ and an auxiliary space $\mathbb{V}$.  The joint measure $\hat{\mu}$ is typically specified through its density $\mu(\theta)\, p(v \mid \theta) $ with respect to a background product measure $(\lambda_{\mathbb{S}} \otimes \lambda_{\mathbb{V}})$, i.e., \[
\hat{\mu}(d\theta \, dv) = \mu(\theta)\, p(v \mid \theta)\, (\lambda_{\mathbb{S}} \otimes \lambda_{\mathbb{V}})(d\theta \, dv),
\]
where $\mu$ is the target density on the original space $\mathbb{S}$, $p(v \mid \theta)$ is a conditional density over auxiliary variables, and $\lambda_{\mathbb{S}}$, $\lambda_{\mathbb{V}}$ are reference measures on $\mathbb{S}$ and $\mathbb{V}$, respectively.    They use the reversibility of $\hat{\pi}$ to derive the reversibility of the marginal kernel on $\mathbb{S}$, and apply their framework to a wide range of MCMC algorithms, including NUTS, Multiple-Try Metropolis, Delayed Rejection, and Event-Chain Monte Carlo methods. % In the special case where the involution $G$ preserves $\lambda_{\mathbb{A}}$, and $a(r) = 1 \wedge r$, their theorem reduces to our Lemma~\ref{lem:DeterministicMetropolization}.

A related and increasingly influential line of work by Glatt-Holtz, Krometis, and Mondaini \cite{Glatt-Holtz_Krometis_Mondaini_2023} develops a general reversibility result for auxiliary variable samplers. Their formulation replaces conditioning with a general refreshment kernel, making it applicable in infinite-dimensional settings. Of particular interest is their introduction of surrogate trajectory HMC methods, where the true potential gradient is replaced with a computationally cheaper surrogate. Their framework provides a rigorous basis for the Metropolis–Hastings ratio in this setting and unifies several classical infinite-dimensional samplers, including preconditioned HMC and the preconditioned Crank–Nicolson algorithm.  This work is further developed in \cite{Glatt-Holtzetal2024}, where trade-offs between surrogate accuracy and sampling efficiency are explored in finite dimensions.

An additional milestone in the theory of locally adaptive HMC is the recent work of Durmus et al.\ \cite{durmus2023convergence}, who established reversibility and geometric ergodicity for a broad class of dynamic HMC algorithms. This foundational result enables sharper theoretical analysis of NUTS and related methods in high-dimensional and anisotropic settings \cite{BouRabeeOberdoerster2024}.

While these general frameworks offer broad applicability, our focus is on the setting where the measure space $(\mathbb{A}, \mathcal{A}, \hat{\mu})$ is equipped with a background reference measure $\lambda_{\mathbb{A}}$ that is preserved by the involution $G$, and with respect to which $\hat{\mu}$ is absolutely continuous.  This setting captures nearly all applications of interest while requiring only minimal measure-theoretic machinery to remain rigorous and general.    It also leads to a particularly simple expression for the acceptance probability, making it computationally attractive.  For these reasons, it is likely to be the setting of primary interest for statistical applications.

The primary contribution of this paper is the GIST sampler (Algorithm~\ref{algo:general-self-tuning-step}), which admits a relatively simple proof of reversibility (Theorem~\ref{thm:gist_reversibility}) that ensures it leaves the target measure invariant.  Moreover, we demonstrate the utility of GIST as a theoretical tool by unifying the reversibility proofs for existing locally adaptive HMC samplers and as a practical tool by introducing simple, novel alternatives to NUTS for locally adapting path lengths.

The rest of the paper is organized as follows.  Section~\ref{sec:gist} presents the main theoretical contributions of the paper: the GIST sampler (Algorithm~\ref{algo:general-self-tuning-step}) and a theorem establishing the reversibility of a class of auxiliary variable methods (Theorem~\ref{thm:auxvarmethod}) including GIST samplers as a special case (Theorem~\ref{thm:gist_reversibility}).  In the context of adaptively sampling path lengths, Section~\ref{sec:pathlength} considers some fundamental special cases of  GIST samplers including randomized HMC (Section~\ref{ex:exact_rHMC}), the Multinomial HMC Sampler (Section~\ref{sec:multinomial_hmc_sampler}), NUTS (Section~\ref{sec:NUTS}), the Apogee-to-Apogee Path Sampler (Section~\ref{sec:Apogee-to-Apogee}), and some novel alternatives to NUTS (Section~\ref{sec:numericalHMC}). 

Sections~\ref{sec:nealsexample} and~\ref{sec:step-distro} are devoted to numerical illustrations and practical implementations of GIST.  
Section~\ref{sec:nealsexample} compares GIST samplers in the context of an ill-conditioned Gaussian target measure. Section~\ref{sec:step-distro} considers path-length adaptation in the discrete-time context, and presents several concrete proposals for sampling the number of leapfrog steps based on the current position and momentum -- holding step size and mass matrix fixed.   Section~\ref{sec:experiments} empirically demonstrates that the current implementation of NUTS in Stan \cite{betancourt2017conceptual} only slightly outperforms a much simpler alternative.

\section{A framework for self-tuning Hamiltonian Monte Carlo}

\label{sec:gist}

\subsection{Auxiliary variable methods}

\label{sec:av}

Here we introduce a  class of auxiliary variable methods for sampling from a given target probability measure $\mu$ on a measurable space $(\mathbb{S}, \mathcal{B})$  \cite[Section 4.1]{Diaconis2007HitandRun}. The key idea is to alternate between updates in the original space and in an augmented space that includes these auxiliary variables.  This approach offers significant flexibility in constructing transition kernels that are reversible with respect to $\mu$, and includes as special cases Metropolis-Hastings, the slice sampler, and locally adaptive HMC methods.

Let $\lambda_\mathbb{S}$ be a positive reference measure on $(\mathbb{S},\mathcal{B})$, e.g., Lebesgue measure on $\R^d$ or the counting measure on a countable space.  Suppose that the target measure $\mu$ has a strictly positive density with respect to $\lambda_{\mathbb{S}}$, also denoted by $\mu(\theta)$ for $\theta \in \mathbb{S}$.  The state space $\mathbb{S}$ is augmented with an auxiliary space $\mathbb{V}$, which is itself a measurable space $(\mathbb{V},\mathcal{V})$ equipped with a reference measure $\lambda_{\mathbb{V}}$.   The product space $(\mathbb{S} \times \mathbb{V}, \mathcal{B} \otimes \mathcal{V})$ is then considered, with the reference measure $\lambda_{\mathbb{S}} \otimes \lambda_{\mathbb{V}}$.  

For each state $\theta \in \mathbb{S}$, define a conditional probability distribution $p( \cdot \mid \theta)$, which is absolutely continuous with respect to $\lambda_{\mathbb{V}}$, with a strictly positive density, also denoted by $p(v \mid \theta)$ for $v \in \mathbb{V}$.  This leads to the following joint distribution on the augmented space  $\mathbb{S} \times \mathbb{V}$ \begin{equation} \label{eq:joint}
\widehat{\mu}(d \theta \, d v) =  \mu(\theta) \, p( v \mid \theta) \, (\lambda_{\mathbb{S}} \otimes \lambda_{\mathbb{V}}) ( d\theta \, dv)  \;.
\end{equation}
%where $\lambda_{\mathbb{S}} \otimes \lambda_{\mathbb{V}}$ is the reference measure on $\mathbb{S} \times \mathbb{V}$.

Next, a measurable involution $G: \mathbb{S} \times \mathbb{V} \to \mathbb{S} \times \mathbb{V}$ (i.e., $G^{-1} = G$) is defined that preserves the reference measure $\lambda_{\mathbb{S}} \otimes \lambda_{\mathbb{V}}$.  A transition step of an auxiliary variable method proceeds as follows. Starting from $\theta \in \mathbb{S}$,
\begin{itemize}
\item Sample $v$ from $p( \cdot \mid \theta)$. 
\item Apply the involution $G$ to  the pair $(\theta,v)$ to obtain $(\theta',v') = G(\theta,v)$. 
\item Discard $v'$ and accept $\theta'$
with probability 
\begin{equation} \label{eq:ap_AV}
a_{\mathrm{AV}}(\theta,v) = \min\!\left( 1,  \ \frac{\mu(\theta') \, p( v' \mid \theta')}{ \mu(\theta) \, p( v \mid \theta)} \right) \;.
\end{equation}
\end{itemize}  Although the procedure alternates between the original space $\mathbb{S}$ and the augmented space $\mathbb{S} \times \mathbb{V}$, the auxiliary variable is discarded after it is used.  As a result, iterating this procedure simulates samples from a Markov chain on the original space $\mathbb{S}$ with transition kernel
\begin{equation}\label{eq:piAV}
\pi_{\mathrm{AV}} (\theta,d\theta') \ = \  \int_{\mathbb{V}} \left[ a_{\mathrm{AV}} (\theta, v) \delta_{\Pi(G(\theta,v))}(d\theta') \, +\, (1 - a_{\mathrm{AV}}(\theta,v) ) \delta_{\theta}(d\theta') \right] p(v \mid \theta) \,  \lambda_{\mathbb{V}}(dv), 
\end{equation}
where $\Pi$ denotes  projection onto the first component (i.e., $\Pi(\theta,v) = \theta$).

\begin{theorem} \label{thm:auxvarmethod}
The transition kernel $\pi_{\mathrm{AV}}$ in \eqref{eq:piAV} is reversible with respect to~$\mu$.
\end{theorem}

A broader formulation of Theorem \ref{thm:auxvarmethod} appears as Theorem 2.1 in \cite{Glatt-Holtz_Krometis_Mondaini_2023}. We present the narrower version here because it aligns directly with the auxiliary variable constructions developed in this paper. 

The proof of Theorem~\ref{thm:auxvarmethod} builds on a special case of the Metropolis-Hastings method that uses deterministic proposals, as discussed in \cite{tierney1998note}.   Specifically, consider a measurable space $(\widehat{\mathbb{S}},\widehat{\mathcal{B}})$ equipped with a reference measure $\lambda_{\widehat{\mathbb{S}}}$, and let $\widehat{\mu}$ be a given target probability measure on this space, absolutely continuous with respect to $\lambda_{\widehat{\mathbb{S}}}$, with a strictly positive density, also denoted by $\widehat{\mu}(x)$ for $x \in \widehat{\mathbb{S}}$.  Suppose $G: \widehat{\mathbb{S}} \to \widehat{\mathbb{S}}$ is a measurable involution  that preserves $\lambda_{\widehat{\mathbb{S}}}$.  The  transition kernel for a Metropolis method with the proposal kernel $q(x,dy) = \delta_{G(x)}(dy)$ and target measure $\widehat{\mu}$ is given by: \begin{equation} \label{eq:pi_det}
\widehat{\pi}(x,dy) \ = \ \widehat{a}(x) \delta_{G(x)}(dy) + (1-\widehat{a}(x)) \delta_x(dy) \;, \qquad \widehat{a}(x) = \min\!\left( 1, \frac{\widehat{\mu}(G(x))}{\widehat{\mu}(x)} \right) \;.
\end{equation}
This construction ensures that $\widehat{\pi}$ is reversible, as formalized in the following lemma.

\begin{lemma} \label{lem:DeterministicMetropolization}
The transition kernel $\widehat{\pi}$ in \eqref{eq:pi_det} is reversible with respect to~$\widehat{\mu}$.
\end{lemma}

Although Lemma \ref{lem:DeterministicMetropolization} is well known \cite{AndrieuLeeLivingstone} , its implications for auxiliary variable methods have not been fully explored. For completeness, we include proofs of both Theorem \ref{thm:auxvarmethod} and Lemma \ref{lem:DeterministicMetropolization} in Appendix \ref{app:proofs}.   A more general version of Lemma~\ref{lem:DeterministicMetropolization} appears as Theorem 3 in \cite{AndrieuLeeLivingstone}.

\begin{example}[Metropolis-Hastings]
\label{eg:MH}
Consider the Metropolis-Hastings method on $\mathbb{S}$ where the target measure $\mu$ and the proposal kernel $q(\theta, d\theta')$ have strictly positive densities with respect to the reference measure  $\lambda_{\mathbb{S}}$, also denoted by $\mu(\theta)$ and $q(\theta, \theta')$, respectively. This method can be interpreted as an auxiliary variable method, where the auxiliary space is $\mathbb{V} = \mathbb{S}$;  the conditional density on $\mathbb{V}$ in \eqref{eq:joint} is defined by $p(\theta' \mid \theta) = q(\theta, \theta')$ for $(\theta, \theta') \in \mathbb{S} \times \mathbb{S}$; and, the measure-preserving involution is defined by $G: (\theta, \theta') \mapsto (\theta', \theta)$ for $(\theta, \theta') \in \mathbb{S} \times \mathbb{S}$.
With these specifications, the acceptance probability from \eqref{eq:ap_AV} then simplifies to:
\[
\alpha_{\mathrm{AV}}(\theta, \theta') = \min\!\left(1 , \ \frac{\mu(\theta') \, q(\theta', \theta)}{\mu(\theta) \, q(\theta, \theta')} \right) \;,
\]
which we recognize as the standard Metropolis-Hastings acceptance probability.
\end{example}

\begin{example}[Slice Sampler]
Consider the slice sampler on $\mathbb{S}$ aimed at a target measure with strictly positive density $\mu(\theta)$ relative to the reference measure $\lambda_{\mathbb{S}}$. The slice sampler  generates a Markov chain by first sampling a height $y$ uniformly from $[0,\mu(\theta)]$ and then sampling from the corresponding ``slice'' of the state space $\mathbb{S}$  where the density exceeds this height, i.e.,  $\textrm{slice}(y) = \{ \theta' \in \mathbb{S} \mid \mu(\theta') > y \} $   \cite{neal1997markov,Neal2003Slice}.  The slice sampler can be interpreted as an auxiliary variable method with the auxiliary space $\mathbb{V}=\mathbb{R} \times \mathbb{S}$, where the reference measure on $\mathbb{V}$ is the product of Lebesgue measures on $\mathbb{R}$ and $\lambda_{\mathbb{S}}$. The conditional density on $\mathbb{V}$ in \eqref{eq:joint}  is  \[
p(y, \theta' \mid \theta) = \mathrm{uniform}( y \mid [0, \mu(\theta)]) \, \mathrm{uniform}( \theta' \mid \textrm{slice}(y) ) \;. \] Define the measure-preserving involution
$G: (\theta,y,\theta') \mapsto (\theta',y,\theta)$ for $(\theta,y,\theta') \in \mathbb{S} \times \mathbb{V}$.   The acceptance probability from \eqref{eq:ap_AV} simplifies to $\alpha_{\mathrm{AV}} \equiv 1$, meaning that the uniform sample from the slice is always accepted.  %Since this is an auxiliary variable method, it follows immediately  from Theorem~\ref{thm:auxvarmethod} that the slice sampler defines a  reversible Markov chain in the original $\theta$-space.  
\end{example}

\begin{algorithm}[t]
\begin{flushleft}
$\textbf{GIST}(\theta, U, p, G)$
\vspace*{2pt}
\hrule
\vspace*{2pt}
\begin{tabular}{ll}
$\theta \in \mathbb{R}^d$ & position
\\
%$\rho \in \mathbb{R}^d$ & momentum (unused---discarded in Gibbs update) \\[2pt] 
% $\alpha \in \mathbb{A}$ & tuning parameter (unused---discarded in Gibbs update) \\[2pt] 
$U: \mathbb{R}^d \to \mathbb{R}$ & potential energy function 
\\
$p(\alpha \mid \theta, \rho ) $ & tuning parameter distribution \\
$G : \mathbb{R}^{2d}  \times \mathbb{A} \rightarrow \mathbb{R}^{2d} \times \mathbb{A}$
& measure-preserving involution 
% $\rho \in \mathbb{R}^d$ & momentum
% \\
% $\alpha \in \mathbb{A}$ & tuning parameter
\end{tabular}
\vspace*{4pt}
\hrule
\vspace*{8pt}
{\footnotesize (GIBBS)} \\[2pt]
 $\rho \sim \textrm{normal}(0, \textrm{I}_{d \times d})$ \hfill (complete momentum refreshment)
\\[4pt]
$\textcolor{blue}{\alpha \sim p(\cdot \mid \theta, \rho)}$ \hfill (tuning parameter refreshment)
\\[4pt]
{\footnotesize (METROPOLIS)} \\[2pt]
 $\textcolor{blue}{(\theta', \rho', \alpha') = G(\theta, \rho,\alpha)   }$ \hfill (compute proposal)
\\[6pt]
$u \sim \textrm{uniform}([0, 1])$ 
\\[6pt]
if $u < \exp\!\left(\frac{1}{2} |\rho|^2 - \frac{1}{2} |\rho'|^2 + U(\theta)- U(\theta')\right) \, \textcolor{blue}{\dfrac{p\left( \alpha' \, | \, \theta', \rho' \right)}{p \left(\alpha\, | \, \theta, \rho \right)}} $  \\
\null \quad return $\theta'$ \hfill (accept) 
\\[8pt]
else \\[-6pt]
\null \quad return $\theta$ \hfill (reject) \\[4pt]
\vspace*{4pt}
\hrule
\caption{\it {\bfseries  GIST Sampler}.  The GIST sampler differs from standard HMC (noted in blue) in the sampling of the tuning parameter, the measure-preserving involution used to compute the proposal, and the subsequent adjustment of the acceptance probability (all highlighted in blue text color).  Here $\emph{I}_{d \times d}$ denotes the $d \times d$ identity matrix.} 
\label{algo:general-self-tuning-step}
\end{flushleft}
\end{algorithm}

\subsection{Locally adaptive Hamiltonian Monte Carlo}

A classic example of an auxiliary variable method is Hamiltonian Monte Carlo (HMC), where auxiliary momentum variables are introduced at each transition step.
HMC samples from an absolutely continuous probability distribution $\mu$ with density $e^{-U(\theta)}$ relative to the Lebesgue measure $m^d$ on $\mathbb{R}^d$ where $U: \mathbb{R}^d \to \mathbb{R}$ is a continuously differentiable potential energy function and $\int_{\mathbb{R}^d} e^{-U(\theta)} d\theta < \infty$.  For simplicity, we assume  the case of a unit mass matrix, so the joint distribution over position $\theta \in \mathbb{R}^d$ and momentum $\rho \in \mathbb{R}^d$ is
\begin{equation} \label{eq:boltzmann}
 \mu \otimes \mathrm{normal}(0, \textrm{I}_{d \times d}) \; ,
\end{equation}
which has density $e^{-H(\theta,\, \rho)}$ relative to the Lebesgue measure $m^{2d}$ on the phase space $\mathbb{R}^{2d}$, where 
\[
H(\theta,\rho) = U(\theta) + \frac{1}{2} \, \rho^{\top} \, \rho \;.
\]
The  $\theta$-marginal of \eqref{eq:boltzmann} is the target probability distribution. In Bayesian inference, $\theta$ is the parameter and $\mu$ is the posterior distribution; while in statistical mechanics, \eqref{eq:boltzmann} is referred to as the Boltzmann distribution.

A key ingredient of any HMC algorithm is a reversible, volume-preserving map $F(\alpha): \mathbb{R}^{2d} \to \mathbb{R}^{2d}$, which is typically obtained from a leapfrog approximation of the flow of Hamilton's equations for the Hamiltonian $H(\theta, \rho)$, i.e.,  \[
\frac{d}{dt} \theta_t = \rho_t \;, \quad \frac{d}{dt} \rho_t = - \nabla U(\theta_t) \;. 
\] 
Here $\alpha$ encapsulates all tuning parameters that will be locally adapted. By definition, this means that $\mathcal{S} \circ F(\alpha)$ is a volume-preserving involution where $\mathcal{S}: \mathbb{R}^{2d} \to \mathbb{R}^{2d}$ is the momentum flip map defined by $\mathcal{S}(\theta, \rho) = (\theta, -\rho)$ \cite{LeRe2004,Hairer2010GeometricNumerical}.  Reversibility and volume preservation ensure that the map $F(\alpha)$ is Metropolis adjustable \cite{BoSaActaN2018}.  We suppose that the tuning parameter $\alpha$
takes values in a set $\mathbb{A}$ with reference measure $\eta$.  

To self tune the parameter $\alpha$, the state space $\mathbb{R}^{2d}$ is enlarged to a product space $\mathbb{R}^{2d} \times \mathbb{A}$ by augmenting the position and momentum variables with an auxiliary tuning parameter variable. On this enlarged space, an enlarged target measure is defined by specifying a conditional distribution of the tuning parameter $\alpha$ given the position and momentum $(\theta, \rho)$, i.e.,  \begin{align} \label{eq:enlarged_target}
\widehat{\mu}(d\theta, d\rho, d\alpha) &\propto  e^{-H(\theta,\rho)} \,  \, p( \alpha \, | \, \theta, \rho) \, (m^{2d} \otimes \eta) (d\theta \, d\rho  \, d\alpha) \;.
\end{align}  Note that the $(\theta, \rho)$-marginal of $\widehat{\mu}$ is the Boltzmann distribution in \eqref{eq:boltzmann}.  Explicit forms of the tuning parameter distribution $p( \alpha \, | \, \theta, \rho)$ are specified in subsequent sections for the special case of path-length self tuning.  

In addition to the tuning parameter distribution, a key component of GIST is a measure-preserving involution defined in the enlarged space.  For any position and momentum $(\theta, \rho) \in \mathbb{R}^{2d}$,
let $G$ be an  involution on $\mathbb{R}^{2d} \times \mathbb{A}$ defined by
\begin{equation}
 \label{eq:G}
 G: (\theta, \rho, \alpha) \mapsto \Big( \mathcal{S} \circ F(\alpha)(\theta, \rho), \, g(\theta,\rho)(\alpha) \Big) 
\end{equation} 
where $g(\theta,\rho) : \mathbb{A} \to \mathbb{A}$ is a measurable map acting on the  tuning parameter $\alpha \in \mathbb{A}$.
For instance, when $g(\theta,\rho)$ is the identity map on $\mathbb{A}$, the map $G$ is a $(m^{2d} \otimes \eta)$-preserving involution, since $\mathcal{S} \circ F(\alpha)$  is a volume-preserving involution on  $\mathbb{R}^{2d}$   (see Corollary~\ref{cor:GIST-reversible}).   In the  case of path-length self-tuning, a non-identity map $g(\theta,\rho)$ is used to demonstrate that GIST encompasses the Multinomial HMC Sampler, NUTS, and the Apogee-to-Apogee Path Sampler as special cases; see \eqref{eq:G_multinomial_hmc}, \eqref{eq:G_nuts}, and~\eqref{eq:G_aaps}, respectively.

In terms of this notation, the GIST sampler in Algorithm~\ref{algo:general-self-tuning-step}  simulates samples from a Markov chain with transition kernel
\begin{equation}\label{eq:piGIST}
\begin{aligned}
& \pi_{\mathrm{GIST}} (\theta,d \theta') \ = \  \int_{\mathbb{R}^d \times \mathbb{A}} \left[ \vphantom{\int_{\mathbb{R}^d \times \mathbb{A}}}  a_{\mathrm{GIST}} (\theta,\rho,\alpha) \delta_{\Pi(G(\theta,\rho,\alpha))}(d \theta') \,  \right. \\ 
& \left. \vphantom{\int_{\mathbb{R}^d \times \mathbb{A}}} \quad \qquad   \qquad  \qquad \qquad  \qquad +\, (1 - a_{\mathrm{GIST}}(\theta,\rho,\alpha) ) \delta_{\theta}(d \theta') \right] p_{\mathrm{GIST}}(\rho, \alpha \mid \theta) \,  (m^d \otimes \eta) (d \rho \, d\alpha) 
\end{aligned}
\end{equation}
where $\Pi$ denotes  projection onto the first component, i.e., $\Pi(\theta,\rho,\alpha) = \theta$, the conditional density of all auxiliary variables is given by \[
p_{\mathrm{GIST}}(\rho, \alpha \mid \theta) = (2 \pi)^{-d/2} \exp\left( -|\rho|^2 /2 \right)  p(\alpha \mid \theta, \rho) 
\] and we introduced the GIST acceptance probability \begin{equation} \label{eq:acceptanceprobability}
a_{\mathrm{GIST}}(\theta, \rho, \alpha) = 1 \wedge \left(  e^{\displaystyle -\Delta H(\theta,\, \rho)} \, \dfrac{p\left(g(\theta,\rho)(\alpha) \, | \, \mathcal{S} \circ F(\alpha)(\theta, \rho) \right)}{ p \left(\alpha\, | \, \theta, \rho \right)} \right) \;,
\end{equation} 
where  $\Delta H(\theta, \rho) := H \circ F(\alpha)(\theta,\rho) - H(\theta, \rho)$ for $(\theta, \rho, \alpha) \in \mathbb{R}^{2d} \times \mathbb{A}$.

The following result implies that the GIST sampler preserves the desired target distribution.  
\begin{theorem} \label{thm:gist_reversibility}
The transition kernel $\pi_{\mathrm{GIST}}$ in \eqref{eq:piGIST} is reversible with respect to $\mu$.
\end{theorem}

The proof of this theorem follows from Theorem~\ref{thm:auxvarmethod}, since GIST is an auxiliary variable method with auxiliary space $\mathbb{V} = \mathbb{R}^d \times \mathbb{A}$, reference measure $\lambda_{\mathbb{V}} = m^d \otimes \eta$, joint distribution in \eqref{eq:enlarged_target}, and measure-preserving involution given by \eqref{eq:G}.

\begin{remark}
Just as the Markov chain for standard HMC can be formulated over phase space variables $(\theta, \rho) \in \mathbb{R}^{2d}$, the Markov chain for GIST can also be formulated over triples $(\theta, \rho, \alpha) \in \mathbb{R}^{2d} \times \mathbb{A}$.  However, in the case treated here of full momentum and tuning parameter refreshment, the sequence of $\theta$ values by itself forms a Markov chain and including the other variables is superfluous.
\end{remark}

\begin{remark}
In addition to adaptively sampling HMC tuning parameters, the GIST sampler can also be applied to randomize the time integrator for the Hamiltonian flow, as in \cite{BouRabeeMarsden2022,BouRabeeKleppe2023}.  In this case, the tuning parameter would specify a particular time integrator within a parametric family of time integrators that are each reversible and volume-preserving.  In certain representative models, randomized time integrators have provably better complexity for Hamiltonian MCMC than the frequently used leapfrog integrator \cite{shen2019randomized,ErgodicityRMMHYB,Cao_2021_IBC,BouRabeeMarsden2022,BouRabeeSchuh2023B,BouRabeeOberdoerster2024}.  
\end{remark}

\begin{remark}
\label{rmk:acceptance}
The GIST acceptance probability $a_{\mathrm{GIST}}$ in \eqref{eq:acceptanceprobability} involves the usual Metropolis ratio, which is a function of the change in energy under $F(\alpha)$, but also an additional term that involves the tuning parameter distribution.  In the subsequent sections, we will see that there is a tradeoff between how strongly the tuning parameter distribution depends on the current position and momentum, and the consequent acceptance probability $a_{\mathrm{GIST}}$.   In particular, NUTS resolves this tradeoff by employing strategic auxiliary variables and a clever tuning parameter distribution that allows for some local dependence, but with $a_{\mathrm{GIST}} \equiv 1$ (i.e., 100\% acceptance).  
\end{remark}

\section{Adaptively sampling path lengths}\label{sec:pathlength}

To make GIST more concrete and demonstrate its breadth, this section develops some novel GIST samplers and proves that existing locally path-length adaptive HMC samplers are GIST samplers.  

\subsection{Randomized HMC as a GIST sampler}\label{ex:exact_rHMC}

In randomized forms of HMC, either the path length or the step size is generated randomly at each step of the chain independently of the current position and momentum \cite{Ma1989}; using, e.g., an exponential distribution over path length \cite{BoSa2017,BoEb2022}, an empirically learned distribution over path length \cite{wu2018faster}, or a uniform distribution over path and step sizes \cite{Ma1989,Ne2011}.  These can all be analyzed as instances of the GIST sampler given in Algorithm~\ref{algo:general-self-tuning-step}.  

As an example, consider $F(\alpha) \equiv \varphi_\alpha$ where  $\varphi_{\alpha}: \mathbb{R}^{2d} \to \mathbb{R}^{2d}$ is the exact Hamiltonian flow at time $\alpha \ge 0$. Take $\mathbb{A} = [0, \infty)$ with background measure given by Lebesgue measure on $\mathbb{R}$. Define $p(\alpha \mid \theta, \rho) = \lambda e^{- \lambda \alpha}$ where $\lambda > 0$ (i.e., $\alpha$ is an exponential random variable with rate $\lambda$).  Define the measure-preserving involution $G$ in \eqref{eq:G} by $G : (\theta, \rho, \alpha) \mapsto ( \mathcal{S} \circ F(\alpha)(\theta, \rho), \ \alpha)$.  With these specifications, $a_{\mathrm{GIST}} \equiv 1$ (the proposal is always accepted), and Algorithm~\ref{algo:general-self-tuning-step} reduces to a draw from randomized HMC at the first jump time \cite{BoSa2017,BoEb2022,kleppe2022connecting}.

The other randomized forms of HMC that randomly generate the tuning parameter independently of the current position and momentum can be analyzed in exactly the same way.  In Section~\ref{sec:nealsexample}, in the setting of a truncation of an infinite-dimensional Gaussian measure, the performance of randomized HMC is compared with a GIST sampler that we describe next. 

\subsection{Novel GIST samplers based on the exact Hamiltonian flow}\label{ex:exact_stHMC}

Consider again $F(\alpha) \equiv \varphi_\alpha$ where  $\varphi_{\alpha}: \mathbb{R}^{2d} \to \mathbb{R}^{2d}$ is the exact Hamiltonian flow map at time $\alpha \ge 0$.  Take $\mathbb{A} = [0, \infty)$ with background measure given by Lebesgue measure on $\mathbb{R}$.  Let $\tau(\theta, \rho):\mathbb{R}^{2d} \to (0, \infty)$ be any measurable function.   Define \[ p(\alpha \mid \theta, \rho) \, = \, \mathrm{uniform}(\alpha \mid [0, \ \tau(\theta, \ \rho)]) \, = \,  \dfrac{1}{\tau(\theta, \rho)} \mathds{1}_{[0, \ \tau(\theta, \ \rho)]}( \alpha ) \;.
\] That is, conditioned on the position and momentum $(\theta, \rho)$, the tuning variable $\alpha$ is a uniform random variable over the interval  $[0,\tau(\theta, \rho)]$.  Here $\mathds{1}_A$ denotes the standard indicator function of a set $A$.  As shorthand, let $\tau_1 = \tau(\theta, \rho)$ and $\tau_2 = \tau(\mathcal{S} \circ \varphi_{\alpha}(\theta, \rho))$, i.e., the function $\tau$ evaluated at the current state of the chain $(\theta, \rho)$ and the proposed state but with momentum reversed $\mathcal{S} \circ \varphi_{\alpha}(\theta, \rho)$.
Define the measure-preserving involution $G$ in \eqref{eq:G} by \[
G : (\theta, \rho, \alpha) \mapsto ( \mathcal{S} \circ F(\alpha)(\theta, \rho), \ \alpha) \;.
\] With these specifications, the corresponding acceptance probability in \eqref{eq:acceptanceprobability} reduces to \[ a_{\mathrm{GIST}}(\theta, \rho, \alpha) \, = \, 1 \wedge \left( \frac{\tau_1}{\tau_2} \mathds{1}_{ \{  \tau_2 \geq \alpha \} } \right) \;, \] because $\Delta H(\theta, \rho) = 0$ for the exact Hamiltonian flow.  The indicator in this Metropolis ratio encodes the requirement  $p(\alpha \mid \mathcal{S} \circ \varphi_{\alpha}(\theta, \rho)) \ne 0$ for non-zero acceptance probability.

Conditions avoiding U-turns in the exact Hamiltonian flow can be used to specify the function $\tau(\theta, \rho)$.  There are several ways to define such U-turn conditions. For example, here is a condition based on when the angle between the initial velocity $\rho \in \mathbb{R}^d$ and the velocity $\rho_t \in \mathbb{R}^d$ at time $t \ge 0$ first exceeds $\pi/2$ \begin{equation} \label{eq:ct_angle}
\tau(\theta, \rho) \ := \ \inf\{ t > 0 ~:~ \rho \cdot \rho_t \le 0 \} \;,
\end{equation} where we have introduced $(\theta_t, \rho_t) := \varphi_t(\theta, \rho)$ for $t \ge 0$.  Another condition could be based on when the squared distance $\Gamma(t) := | \theta - \theta_t|^2$ between the initial configuration $\theta \in \mathbb{R}^d$ and the configuration $\theta_t \in \mathbb{R}^d$ at time $t \ge 0$ first decreases \begin{equation} \label{eq:ct_dist}
\tau(\theta, \rho) \ := \ \inf\{ t > 0 ~:~ \Gamma'(t) < 0 \} \;.
\end{equation}
Discrete time analogs of these continuous-time U-turn conditions are explored in Section ~\ref{sec:step-distro}.  Note, the exact Hamiltonian flow will always eventually undergo a U-turn for potentials with asymptotic growth at infinity – which must be the case for normalizable target densities. But, in practice, one imposes a cap on the length of trajectories which may be triggered before a U-turn actually occurs.

In this context, it also possible to design GIST samplers with the property $a_{\mathrm{GIST}} \equiv 1$.  Take $\alpha = (a,b,t) \in \mathbb{A} = \mathbb{R} \times \mathbb{R} \times \mathbb{R}$ with background measure given by Lebesgue measure.  Let $\tau(\theta, \rho):\mathbb{R}^{2d} \to [0, \infty)$ be defined by \begin{equation} \label{eq:ct_virial}
\tau(\theta, \rho) \ := \ \inf \left\{ s \ge 0 ~~ \mid  ~~ \rho_s \cdot \theta_s  = 0 \right\} \;. 
\end{equation} Note that the zero set of the virial function $\rho_s \cdot \theta_s$ is not empty, since the level sets of the underlying Hamiltonian are bounded and $\frac{d}{ds} |\theta_s|^2 = 2 \rho_s \cdot \theta_s$.   Define the tuning parameter  distribution in \eqref{eq:joint} by \begin{equation} \label{eq:p_exact_virial}
p(a, b, t \mid \theta, \rho)  \, = \,  \dfrac{1}{b - a} \mathds{1}_{[a,  b]}( t ) \; \delta_{-\tau \circ \mathcal{S}  (\theta,\rho)}(a) \; \delta_{\tau(\theta,\rho)}(b) \;.
\end{equation} Here $\delta_x$ is the Dirac-delta function concentrated at the point $x \in \mathbb{R}$.
Define the measure-preserving involution $G$ in \eqref{eq:G} by \begin{equation} \label{eq:G_exact_virial}
G : (\theta, \rho, a, b, t) \mapsto ( \mathcal{S} \circ \varphi_t(\theta, \rho), t-b,   t - a , t) \;.
\end{equation} With these specifications, the corresponding acceptance probability in \eqref{eq:acceptanceprobability} reduces to $a_{\mathrm{GIST}} \equiv 1$.  This result follows by inserting \eqref{eq:p_exact_virial} and \eqref{eq:G_exact_virial} into \eqref{eq:acceptanceprobability} and using the identities: 
\begin{equation}
\label{eq:exact_virial_identities}
\begin{aligned}
\tau \circ \varphi_{t}(\theta, \rho)  &= \inf\{ s>0 \mid \rho_{s + t} \cdot \theta_{s + t} = 0 \} = \tau(\theta, \rho) - t \;, \\
\tau \circ \mathcal{S} \circ \varphi_{t}(\theta, \rho)  &= \inf\{ s>0 \mid \rho_{-s + t} \cdot \theta_{-s + t} = 0 \} = \tau \circ \mathcal{S} (\theta, \rho) + t \;,  
\end{aligned}
\end{equation}
which hold for any $t \in [-\tau \circ \mathcal{S} (\theta, \rho), \tau(\theta, \rho)]$.   Indeed, since this GIST sampler uses the exact Hamiltonian flow, where $\Delta H \equiv 0$, the only contribution to \eqref{eq:acceptanceprobability} comes from the conditional distribution of the tuning parameter, which satisfies:  \begin{align*}
& p( g(\theta,\rho) (\alpha)  \mid \mathcal{S} \circ \varphi_t(\theta, \rho) )   \, \overset{\eqref{eq:p_exact_virial}}{=} \,    \dfrac{1}{b - a} \mathds{1}_{[t-b,  t-a]}( t ) \; \delta_{-\tau \circ  \varphi_t(\theta,\rho)}(t-b) \; \delta_{\tau \circ \mathcal{S} \circ \varphi_t(\theta,\rho)}(t-a) \, \\ 
&\qquad \overset{\eqref{eq:exact_virial_identities}}{=} \,    \dfrac{1}{b - a} \mathds{1}_{[t-b,  t-a]}( t ) \; \delta_{-\tau (\theta,\rho) + t}(t-b) \; \delta_{\tau \circ \mathcal{S} (\theta,\rho)+t}(t-a)  \, =  \,  p( \alpha  \mid \theta, \rho )    \;,
\end{align*}
where in the last step the scaling and sifting properties of the Dirac-delta function were used. Hence, $a_{\mathrm{GIST}} \equiv 1$.

\begin{remark}
    The condition appearing in the function $\tau$ in \eqref{eq:ct_virial} is somewhat related to the notion of an \emph{exhaustion} used in Exhaustive HMC; see Definition 1 of~\cite{betancourt2016identifying}.  The term exhaustion refers to integrating the Hamiltonian flow for an exhaustive period of time  in order to explore phase space more thoroughly.  Like \eqref{eq:ct_virial}, the criterion for exhaustion used in Exhaustive HMC is based on the time evolution of the virial function along the Hamiltonian flow.
\end{remark}

\subsection{Novel GIST samplers based on the leapfrog integrator}\label{sec:numericalHMC}

Next, we consider GIST samplers that can be implemented for general models with the leapfrog integrator.  First, fix a step size $h>0$.  Let $\Phi_h: \mathbb{R}^{2d} \to \mathbb{R}^{2d}$ denote one step of the leapfrog integrator with step size $h$.  Consider $F(\alpha) \equiv \Phi_h^{\alpha}$ where $\alpha$ is the number of leapfrog steps. Take $\mathbb{A} = \mathbb{N}$ with background measure given by the counting measure.  For any $m,n \in \mathbb{Z}$ such that $m \ge n$, let \[ [m:n] = \{ m, m+1, ..., n \} \;.
\]
That is, $[m:n]$ is the set of consecutive integers from $m$ to $n$.  Let $\tau: \mathbb{R}^{2d} \to \mathbb{N}$ be a measurable function and define \[ p(\alpha \mid \theta, \rho) \, = \, \mathrm{uniform}(\alpha \mid [0 : (\tau(\theta, \, \rho)-1)]) \, = \,  \frac{1}{\tau(\theta, \rho)} \mathds{1}_{[0 : (\tau(\theta, \, \rho)-1)]}(\alpha) \;.  \]
Define the measure-preserving involution $G$ in \eqref{eq:G} by \[
G : (\theta, \rho, \alpha) \mapsto ( \mathcal{S} \circ F(\alpha)(\theta, \rho), \ \alpha) \;.
\] With these specifications, the corresponding acceptance probability in \eqref{eq:acceptanceprobability} simplifies to \[
a_{\mathrm{GIST}}(\theta, \rho, \alpha) \, = \, 
1 \wedge \left( e^{-\Delta H(\theta,\, \rho)} \frac{\tau_1}{\tau_2} \mathds{1}_{\{ \tau_2 \geq \alpha \} } \right) \;, \]
where $\tau_1 = \tau(\theta, \rho)$ and $\tau_2 = \tau(\mathcal{S} \circ F(\alpha)(\theta, \rho))$.  A range of concrete choices for $\tau$ are described and evaluated in  Section~\ref{sec:step-distro}. 

The following corollary of Theorem~\ref{thm:gist_reversibility} ensures that this class of GIST samplers is reversible.

\begin{corollary} \label{cor:GIST-reversible}
If $G : (\theta, \rho, \alpha) \mapsto ( \mathcal{S} \circ F(\alpha)(\theta, \rho), \ \alpha)$ for all $(\theta, \rho, \alpha) \in \mathbb{R}^{2d} \times \mathbb{A}$, the transition kernel of the corresponding GIST sampler is reversible with respect to $\mu$.
\end{corollary}

The proof follows by verifying that the map $G$ is a measure-preserving involution, and is provided in Appendix~\ref{app:proofs} for the reader's convenience.

In this context, one can also obtain GIST samplers based on the leapfrog integrator with the property $a_{\mathrm{GIST}} \equiv 1$.   Specifically,  for $k \in \mathbb{Z}$,  let $(\theta^{(k)}, \rho^{(k)}) = \Phi_h^k(\theta, \rho)$ be the $k$-th leapfrog iterate.  Let $L_a, L_b \in \mathbb{Z}$ define the index set $[L_a:L_b]$ that labels a subset of leapfrog iterates that can be transitioned to.  
Define $\Lambda(h): \mathbb{R}^{2d} \to \mathbb{N}$ by \begin{equation} \label{eq:lambda_h}
\Lambda(h)(\theta, \rho) \ := \ \min \left\{ k \in \mathbb{N} ~~ \mid  ~~ \sign(\theta^{(k)} \cdot \rho^{(k)}) \ne \sign(\theta^{(k+1)} \cdot \rho^{(k+1)})  \right\} \;. 
\end{equation} 
This function is the discrete-time analog of \eqref{eq:ct_virial}.  Take $\alpha = (L_a,L_b,L) \in \mathbb{A} = \mathbb{Z}^3$, and define the tuning parameter distribution in \eqref{eq:joint} by \begin{equation} \label{eq:nu_dt_exact_rev_GIST}
p(L_a, L_b, L \mid \theta, \rho) \ = \  \dfrac{e^{-H \circ \Phi_h^{L}(\theta, \rho)}}{\sum_{k \in [L_a:L_b]} e^{-H \circ \Phi_h^{k}(\theta, \rho)}} \; \mathbb{1}_{[L_a:L_b]}(L) \; \delta_{-\Lambda(h) \circ \mathcal{S}  (\theta,\rho)} ( L_a) \; \delta_{\Lambda(h)(\theta,\rho)}( L_b)  \;.
\end{equation} In other words, given $(\theta, \rho) \in \mathbb{R}^{2d}$, the endpoints of the index set $[L_a:L_b]$ are chosen \emph{deterministically} by evaluating $\Lambda(h)$ at $(\theta, \rho)$ and $\mathcal{S}(\theta, \rho)$, and in turn, a leapfrog iterate $L \in [L_a:L_b]$ is randomly sampled from the  categorical distribution with density \begin{equation} 
Q(L  \mid \theta, \rho, L_a, L_b ) = e^{-H \circ \Phi_h^{L}(\theta, \rho)} \, \left( \sum_{k \in [L_a:L_b]} e^{-H \circ \Phi_h^{k}(\theta, \rho)} \right)^{-1} \, \mathbb{1}_{[L_a:L_b]}(L) \;. 
\end{equation} To complete the specification of this GIST sampler, define the measure-preserving involution \eqref{eq:G} by \begin{equation} \label{eq:G_dt_exact_rev_GIST}
G : (\theta, \rho, L_a, L_b, L) \mapsto ( \mathcal{S} \circ \Phi_h^{L}(\theta, \rho), L-L_b,   L - L_a , L) \;.
\end{equation} 
With these specifications, it is straightforward to show that the corresponding GIST acceptance probability in \eqref{eq:acceptanceprobability} satisfies $a_{\mathrm{GIST}} \equiv 1$.

\subsection{The multinomial HMC sampler as a GIST sampler}\label{sec:multinomial_hmc_sampler}

Here we show that the Multinomial HMC Sampler is a GIST sampler \cite{betancourt2017conceptual,xu2021couplings}.  The proof illustrates some of the subtleties associated with locally adapting path lengths.  Besides its pedagogical value, the Multinomial HMC Sampler can also be seen as an implementable discretization of randomized HMC from Section~\ref{ex:exact_rHMC}.

 Let $L_{max} \in \mathbb{N}$ be the maximum number of leapfrog evaluations at each step of the chain, which effectively fixes the computational budget.  Define the following set of \emph{orbits}
  \[
\mathcal{O} = \{ [-L_{max} :0], [-L_{max} +1:1], \dots, [0:L_{max} ] \} \;. \] 
In other words, $\mathcal{O}$ consists of all sets of consecutive integers of size $L_{max}+1$ that contain zero.   %Let $\Phi_h: \mathbb{R}^{2d} \to \mathbb{R}^{2d}$ denote one step of the leapfrog integrator with step size $h$. 

Given the current state of the chain $\theta \in \mathbb{R}^d$, a transition step of the Multinomial HMC Sampler with complete velocity refreshment proceeds as follows. First, the velocity is completely refreshed $\rho \sim \operatorname{normal}(0, \textrm{I}_{d \times d})$.  Then, $J$ is uniformly sampled from $\mathcal{O}$, and in turn, an index $i$ is sampled from the orbit $J$ with probability proportional to the corresponding state's Boltzmann weight, where the weights are $\{ e^{-H \, \circ \, \Phi_h^{j}(\theta,\, \rho)} \}_{j \in J } $.   The next state of the chain is then $\theta'$ where $(\theta', \rho') = \Phi_h^i(\theta, \rho)$. In particular, the proposal is always accepted.

To formulate the Multinomial HMC Sampler as a GIST sampler, set
 $\alpha = (J, i)$ and $F(\alpha) \equiv \Phi_h^{i}$ where $J \in \mathcal{O}$ and $i \in J$. Correspondingly, set $\mathbb{A} = \mathcal{O} \times \mathbb{Z}$ with background measure given by the counting measure. For $J \in \mathcal{O}$ and $i \in J$, define the conditional distribution of $(J,i)$ given  $(\theta, \rho) \in \mathbb{R}^{2d}$ by 
\begin{align} & p(J, i \mid \theta, \rho) \ = \ P_h(J \mid \theta, \rho) \; Q_h(i \mid \theta, \rho, J) \;, \label{eq:p_multinomial_hmc} \\ 
& \text{where} \quad 
\begin{cases}
P_h(J \mid \theta, \rho) \ = \  (L_{max} +1)^{-1} \, \mathds{1}_{ \mathcal{O} } (J) \;, \\
Q_h(i \mid \theta, \rho, J) \ = \  e^{-H \, \circ \, \Phi_h^{i}(\theta,\, \rho)} \left( \sum_{k \in J} e^{-H \, \circ \, \Phi_h^{k}(\theta,\, \rho)} \right)^{-1} \, \mathds{1}_{ J } (i) \;. 
\end{cases} \nonumber
\end{align}
In the literature, $P_h$ and $Q_h$ are called the orbit and index selection densities, respectively \cite{betancourt2017conceptual,durmus2023convergence}. 
Note, conditional on $(\theta,\rho)$ and $J$, the index $i$ is a sample from the categorical distribution with density: \begin{equation} \label{eq:categorical}
Q_h(i \mid \theta, \rho, J) \ = \ \operatorname{categorical}( i \mid  \{ Z^{-1} e^{-H \circ \Phi^k_h(\theta, \rho)} \}_{k \in J} ) \;,  \quad \text{where}~~Z = \sum_{j \in J} e^{-H \, \circ \, \Phi_h^{j}(\theta,\, \rho)} \;.
\end{equation}
Define the measure-preserving involution $G$ in \eqref{eq:G} by   \begin{equation}
\label{eq:G_multinomial_hmc}
    G : (\theta, \rho, J, i) \mapsto ( \mathcal{S} \circ \Phi_h^{i}(\theta, \rho), -( J-i), i) \;.
\end{equation}

Note that for all $J \in \mathcal{O}$ and $i \in J$, $-(J-i) \in \mathcal{O}$ and $i \in -(J-i)$.
 With these specifications,  $a_{\mathrm{GIST}} \equiv 1$ in the corresponding GIST sampler. Indeed, for all $J \in \mathcal{O}$ and $i \in J$, \begin{align*}
 p(-(J-i), i \mid \mathcal{S} \circ \Phi_h^i(\theta, \rho)) &= \frac{1}{L_{max}+1} \mathds{1}_{ \mathcal{O}}(-(J-i))  e^{-H(\theta,\, \rho)} \left( \sum_{k\in -(J-i)} e^{-H \, \circ \, \Phi_h^{i-k}(\theta,\, \rho)} \right)^{-1} \mathds{1}_{ -(J-i) } (i) \;, \\
&  = \frac{e^{-H(\theta,\, \rho)}}{L_{max}+1}  \left( \sum_{k\in J} e^{-H \, \circ \, \Phi_h^{k}(\theta,\, \rho)} \right)^{-1} = e^{-H(\theta,\, \rho) + H \, \circ \, \Phi_h^{i}(\theta,\, \rho)} p(J, i \mid \theta, \rho)  \;.
\end{align*}
Inserting this result into \eqref{eq:acceptanceprobability} and simplifying yields $a_{\mathrm{GIST}} \equiv 1$, as claimed.  Moreover, the corresponding GIST sampler in Algorithm~\ref{algo:general-self-tuning-step} reduces to a transition step of the Multinomial HMC Sampler, which has the pleasant feature that the proposal is always accepted.

\subsection{The No-U-Turn sampler as a GIST sampler}\label{sec:NUTS} 
In this part, we show that revised NUTS is a GIST sampler \cite{betancourt2017conceptual}. Moreover, in Remark~\ref{rmk:originalNUTS} below, we  also show that the original slice sampler version of NUTS is a GIST sampler \cite{HoGe2014}.  The revised version of NUTS can be viewed as a generalization of the Multinomial HMC Sampler from Section~\ref{sec:multinomial_hmc_sampler} that incorporates U-turn avoiding conditions.  We first describe a transition step of NUTS more precisely following \cite{betancourt2017conceptual,durmus2023convergence}. We then specify the tuning parameter distribution and measure-preserving involution in the enlarged space.

Similarly to the Multinomial HMC Sampler, NUTS randomly generates an orbit $J$, and in turn, randomly samples an index $i$ from this orbit $J$. The next state of the chain is then $\theta'$ where $(\theta', \rho') = \Phi_h^i(\theta, \rho)$.  As in the Multinomial HMC Sampler, let $L_{max}$ denote the maximum size of the orbit $J$. As NUTS uses the \emph{random doubling procedure} described below to generate $J$, $L_{max}$ can always be taken to be a power of $2$. As before, let $P_h(J \mid \theta, \rho)$ be the probability of selecting orbit $J$ and let $Q_h(i \mid J, \theta, \rho)$ be the probability  of selecting $i \in J$.

For any set of consecutive integers $I \subset \mathbb{Z}$ of size $|I| \le L_{max}$, use the indicator  $\mathds{1}_{\text{U-turn}}(I,\theta,\rho)$ to indicate that the set of leapfrog iterates $\{ \Phi_h^k(\theta, \rho)\}_{k \in I}$ has hit a U-turn  \[
\text{U-turn} = \{ (I, \theta, \rho) ~\mid~
\rho_+ \cdot (\theta_+ - \theta_-) < 0 \quad \text{and} \quad \rho_- \cdot (\theta_+ - \theta_-) < 0 \} \]
where $(\theta_+, \rho_+) = \Phi_h^{\max{I}}(\theta, \rho)$ and $(\theta_-, \rho_-) = \Phi_h^{\min{I}}(\theta, \rho)$.  Note that this U-turn condition depends on $(\theta, \rho)$ only through the endpoints of the leapfrog iterates $\{ \Phi_h^k(\theta, \rho)\}_{k \in I}$.  This property is key to obtaining $a_{\mathrm{GIST}} \equiv 1$ in the corresponding GIST sampler (cf.~Remark~\ref{rmk:acceptance}). 

Additionally, for any consecutive set of integers $I \subset \mathbb{Z}$  of size $|I|=2^k$ for $k \in \mathbb{N}$, the random doubling procedure used by NUTS employs $\mathds{1}_{\text{sub-U-turn}}(I,\theta,\rho)$ to indicate the  presence of a \emph{sub-U-turn}  defined recursively by  
\begin{equation}
\mathds{1}_{\text{sub-U-turn}}(I,\theta,\rho) = \max\{ \mathds{1}_{\text{sub-U-turn}}(I^-,\theta,\rho),\, \mathds{1}_{\text{sub-U-turn}}(I^+,\theta,\rho),\, \mathds{1}_{\text{U-turn}}(I,\theta,\rho) \} \label{eq:sub-u-turn-def}
\end{equation}
where $I^-$ and $I^+$ are defined as the left and right halves of $I$, respectively.  Moreover, if $|I|=1$, then naturally $\mathds{1}_{\text{sub-U-turn}}(I,\theta,\rho) = 0$. We emphasize that $\mathds{1}_{\text{sub-U-turn}}(I,\theta,\rho)=0$ does not imply that $\mathds{1}_{\text{U-turn}}(K,\theta,\rho)=0$ for all consecutive sets of integers $K \subset I$.

The random doubling procedure for generating $J$ then proceeds as follows: take $J_{0} = \{ 0 \}$.  This initialization ensures that every generated orbit contains zero. Given $J_k$, if $|J_k| = L_{max}$ or $\mathds{1}_{\text{U-turn}}(J_k, \theta, \rho)=1$, then the procedure returns $J=J_k$; otherwise, a proposed extension  $I_{k+1}^*$ is sampled uniformly from $\{ J_k + |J_k|, J_k - |J_k| \}$.  If $\mathds{1}_{\text{sub-U-turn}}(I_{k+1}^*,\theta, \rho) = 1$, then the procedure returns $J=J_k$; otherwise, $J_{k+1} = J_k \cup I_{k+1}^*$ and the procedure is iterated.

Having generated $J$ using this random doubling procedure, NUTS then generates $i \in J$ according to the index selection kernel $Q_h(i \mid \theta, \rho, J)$. 
There are two index selection kernels commonly used in practice: (i) categorical sampling, as in the Multinomial HMC Sampler from Section~\ref{sec:multinomial_hmc_sampler}; and (ii) biased progressive sampling, as detailed in \cite{betancourt2017conceptual}.

We are now in position to describe NUTS as a GIST sampler. Define the set of orbits $\mathcal{O}$ by
\[ \mathcal{O} = \{J  \subset \mathbb{Z} ~\mid~ J = [m:n]~\text{for}~m,n \in \mathbb{Z},~~ |J| \le L_{max}~~ \text{and}~~0 \in J \} \;.
\] 
As in the Multinomial HMC Sampler, write $\alpha = (J, i)$ for $J \in \mathcal{O}$ and $i \in J$ and set $F(\alpha) \equiv \Phi_h^i$. Correspondingly, set $\mathbb{A} = \mathcal{O} \times \mathbb{Z}$ 
with background measure given by the counting measure.

For all $(\theta, \rho, J, i) \in \mathbb{R}^{2d} \times \mathcal{O} \times \mathbb{Z}$, define
\begin{equation} \label{eq:p_nuts}
    p(J, i \mid \theta, \rho) = P_h(J \mid \theta, \rho) \, Q_h(i \mid \theta, \rho, J) \;,
\end{equation}
and define the measure-preserving involution $G$  in (\ref{eq:G}) by
\begin{equation} \label{eq:G_nuts}
G: (\theta, \rho, J, i) \mapsto (\mathcal{S} \circ \Phi_h^i(\theta, \rho), -(J- i), i) \;.
\end{equation}
With these specifications, the resulting GIST sampler in Algorithm~\ref{algo:general-self-tuning-step} reduces to NUTS. Notably, for NUTS, the proposal is always accepted, i.e., $a_{\mathrm{GIST}} \equiv 1$. Indeed, since for any consecutive set of integers $I \subset \mathbb{Z}$, \[ \{\Phi_h^i \circ \mathcal{S} (\theta, \rho)\}_{i \in -I} = \{\Phi_h^{-i} \circ \mathcal{S}(\theta, \rho)\}_{i \in I} =\{\mathcal{S} \circ \Phi_h^{i}(\theta, \rho)\}_{i \in I}
\] a short coupling construction (see Remark~\ref{rmk:coupling}) shows that $P_h(-J \mid \mathcal{S}(\theta, \rho)) = P_h(J \mid \theta, \rho)$. Combining this with the symmetry condition $
P_h(J-i \mid \Phi_h^i(\theta, \rho)) = P_h(J \mid \theta, \rho)$ (see Remark ~\ref{rmk:initial-point-kernal-symmetry}) shows that the orbit selection kernel satisfies: \[
P_h(-(J-i) \mid \mathcal{S} \circ \Phi_h^i(\theta, \rho)) = P_h(J \mid \theta, \rho) \;.
\] Since the index selection kernels described above also satisfy the detailed balance condition
\[e^{-H(\Phi_h^i(\theta,\, \rho))} Q_h(i \mid \mathcal{S} \circ \Phi_h^i(\theta, \rho), -(J - i)) = e^{-H(\theta,\, \rho)} Q_h(i \mid \theta, \rho, J) \;, \]
it follows that $a_{\mathrm{GIST}} \equiv 1$, as claimed. For the index selection kernel based on the Boltzmann weights this detailed balance condition can be seen by a direct computation, and for biased progressive sampling this property follows from Proposition 6 in \cite{durmus2023convergence}.

To complete the proof that NUTS is a GIST sampler, it remains to verify that $G$ in \eqref{eq:G_nuts} is a measure-preserving involution.  Once this is established, the following corollary follows directly from Theorem~\ref{thm:gist_reversibility}.

\begin{corollary} \label{cor:NUTS-reversible}
The transition kernel of the No-U-Turn Sampler is reversible with respect to $\mu$. 
\end{corollary}

The proof of the above can be found in Appendix \ref{app:proofs}. An analogous corollary holds for the Multinomial HMC Sampler described in Section~\ref{sec:multinomial_hmc_sampler}, and its proof follows by similar arguments, so it is omitted here.

\begin{remark}[Original Slice Sampler Version of NUTS as a GIST Sampler]
\label{rmk:originalNUTS}
The original version of NUTS \cite{HoGe2014} is also a GIST sampler where the auxiliary variable includes a slice variable $s>0$, i.e., $\alpha = (s, J, i)$ for $J \in \mathcal{O}$ and $i \in J$.  Similar to \eqref{eq:p_nuts}, the corresponding tuning parameter distribution is \[
p(s, J, i \mid \theta, \rho) \ = \  P_{\text{slice}}(s \mid \theta, \rho) \, \tilde{P}_h(J \mid \theta, \rho, s) \, \tilde{Q}_h(i \mid J ) \;,
\]
where we have introduced an orbit selection kernel $\tilde{P}_h$ (defined below), and respectively, the following slice selection kernel $P_{\text{slice}}$ and index selection kernel $\tilde{Q}_h$: \begin{align*}
    P_{\text{slice}}(s \mid \theta, \rho) \ = \  \frac{1}{e^{-H(\theta, \rho)}} \mathbb{1}_{[0,e^{-H(\theta, \rho)}]}(s) \;, \quad \text{and} \quad \tilde{Q}_h( i \mid J ) \ = \  \frac{1}{|J|} \mathbb{1}_J(i) \;. 
\end{align*}  
Conditional on the current state $(\theta, \rho)$ and the slice variable $s>0$, the orbit selection kernel $\tilde{P}_h$ is defined implicitly in terms of the following subset of the orbit $J'$ generated by the random doubling procedure described above \[
J \ = \ \{ k \in J' \mid s \le e^{-H \circ \Phi^k_h(\theta, \rho)} \} \;.
\]
That is, $J$ consists of the subset of $J'$ corresponding to the leapfrog iterates that belong to the slice. Define the measure-preserving involution on the enlarged space as \[
G(\theta,\rho,s,J,i)  \ = \  (S \circ \Phi_h^i(\theta, \rho), s, -(J-i), i) \;. 
\] With these specifications, the corresponding GIST sampler is the original slice sampler version of NUTS, and importantly, it is straightforward to show that the corresponding GIST acceptance probability satisfies $a_{\mathrm{GIST}} \equiv 1$.   In particular, the normalization factors in $P_{\text{slice}}$ cancel with the usual Metropolis ratio $e^{-\Delta H(\theta, \ \rho)}$, and in this sense, play a similar role as the Boltzmann weights in the categorical distribution for index selection in \eqref{eq:categorical}. 
\end{remark}

\begin{remark}[Reflection symmetry in orbit selection]

\label{rmk:coupling}

To prove that $P_h(J \mid \theta, \rho) = P_h(-J \mid \mathcal{S}(\theta, \rho))$, we produce a coupling between two copies of the random doubling procedure started at $(\theta,\rho)$ and $\mathcal{S}(\theta, \rho)$. For simplicity we neglect the maximum size restriction, but it should be clear that the argument works even if this restriction is imposed. Let $\{ U_i \}_{i \in \mathbb{N}}$ be symmetric Rademacher random variables. Let $J_0 = \tilde{J}_0 = \{ 0 \}$. Given $J_k$ and $\tilde{J}_k$, if 
\begin{equation}
\mathbb{1}_{\text{U-turn}}(J_k, \theta, \rho) =1  \quad \text{   or    } \quad \mathbb{1}_{\text{U-turn}}(\tilde{J}_k, \mathcal{S}(\theta, \rho)) = 1 \label{eq:coupuled-u-turn}
\end{equation} output $(J_k, \tilde{J}_k)$. Otherwise, let $I_{k+1}^\ast = J_k + U_k \cdot |J_k|$ and $\tilde{I}_{k+1}^\ast = \tilde{J}_k - U_k \cdot |\tilde{J}_k|$. If  
\begin{equation}
\mathbb{1}_{\text{sub-U-turn}}(I^\ast_{k+1}, \theta, \rho) =1 \quad \text{or} \quad \mathbb{1}_{\text{sub-U-turn}}(\tilde{I}^\ast_{k+1}, \mathcal{S}(\theta, \rho)) = 1 \label{eq:coupled-sub-u-turn}
\end{equation} then output $(J_k, \tilde{J}_k)$. Otherwise, $J_{k+1} = J_k \cup I_{k+1}^\ast$ and $\tilde{J}_{k+1}^\ast = \tilde{J}_k \cup \tilde{I}_{k+1}^\ast$ and the procedure is iterated. Let $(J, \tilde{J})$ be the resulting sample. 

One can show by induction that at each stage $\tilde{J}_k = - J_k$ 
and $\tilde{I}_{k+1}^\ast = - I_{k+1}^\ast$. In general, for any set $K$ of consecutive integers, $\mathbb{1}_{\text{U-turn}}(K, \theta, \rho) = \mathbb{1}_{\text{U-turn}}(-K, \mathcal{S}(\theta, \rho))$ and $\mathbb{1}_{\text{sub-U-turn}}(K, \theta, \rho) =\mathbb{1}_{\text{sub-U-turn}}(-K, \mathcal{S}(\theta, \rho))$. Indeed, since $\Phi_h^i \circ \mathcal{S} (\theta, \rho) = \mathcal{S} \circ \Phi_h^{-i}(\theta, \rho)$ the set of leapfrog iterates corresponding to a set $-K$ started from $\mathcal{S}(\theta, \rho)$ is the same as the set of leapfrog iterates for $K$ starting from $(\theta, \rho)$ with all their momenta flipped. Moreover, $\max(-K) = - \min(K)$ and $\min(-K) = - \max(K)$. Consequently, letting $(\theta_+, \rho_+)$ and $(\theta_-, \rho_-)$ be the terms appearing in the U-turn condition for $K$ started at $(\theta, \rho)$ and $(\tilde{\theta}_+, \tilde{\rho}_+), (\tilde{\theta}_-, \tilde{\rho}_{-})$ be the corresponding terms for $-K$ started from $\mathcal{S}(\theta, \rho)$ we have the equalities 
\[(\tilde{\theta}_+, \tilde{\rho}_+) = (\theta_-, - \rho_-) \quad (\tilde{\theta}_-, \tilde{\rho}_-) = (\theta_+, - \rho_+)\]
and therefore 
\begin{align*}
\rho_+ \cdot (\theta_+ - \theta_-) &= \tilde{\rho}_- \cdot (\tilde{\theta}_+ - \tilde{\theta}_-) \\
\rho_- \cdot (\theta_+ - \theta_-) &= \tilde{\rho}_+ \cdot (\tilde{\theta}_+ - \tilde{\theta}_-)
\end{align*}
and consequently if one set has a U-turn, so does the other. The same conclusion holds for sub-U-turns by using the above, (\ref{eq:sub-u-turn-def}), and induction. 

Because of this, checking both of the indicators in each of the termination conditions \eqref{eq:coupuled-u-turn} and \eqref{eq:coupled-sub-u-turn} is redundant. Instead, we could check either the first in both or the second in both. If we check the first indicator in both equations \eqref{eq:coupuled-u-turn} and \eqref{eq:coupled-sub-u-turn}, it is clear that $J$ is a sample from $P_h(\cdot \mid \theta, \rho)$. On the other hand, if we check the second indicator in both equations it is clear that $\tilde{J}$ is a sample from $P_h(\cdot \mid \mathcal{S}(\theta, \rho))$. Additionally, since $\tilde{J}_k = - J_k$ in each step, we have $\tilde{J} = - J$, and consequently we find $P_h(J \mid \theta, \rho) = P_h(-J \mid \mathcal{S}(\theta, \rho))$.

\end{remark}
\begin{remark}[Initial point symmetry in orbit selection and rejection of orbit extensions] \label{rmk:initial-point-kernal-symmetry}

Obtaining $a_{\mathrm{GIST}} \equiv 1$ in NUTS relied critically on   $P_h(J-i \mid \Phi_h^i(\theta, \rho)) = P_h(J \mid \theta, \rho)$ for $i \in J$. This symmetry ensures  that conditioned on a set of leapfrog iterates corresponding to the output of the random doubling procedure, each iterate within this set is equally likely to have been the initial point. Indeed, the set of leapfrog iterates $\{ \Phi_h^i(\theta, \rho) \}_{i \in J}$ could in principle have been produced starting at any of the points within this set. The random doubling procedure would in this case have passed through an alternate sequence of intermediate sets $\tilde{J}_0, \tilde{J}_1, \dots$.
 
A necessary condition to get $P_h(J-i \mid \Phi_h^i(\theta, \rho)) = P_h(J \mid \theta, \rho)$ for $i \in J$ is that none of these alternative sequences would have prematurely terminated the random doubling procedure, i.e., there are no premature sub-U-turns: for an alternative starting point $\Phi_h^i(\theta, \rho)$, $i \in J$ with corresponding alternative intermediate sets $\tilde{J}_0, \tilde{J}_1, \dots$ we need $\mathbb{1}_{\text{U-turn}}(\tilde{J}_k, \Phi_h^i(\theta, \rho)) = 0$ unless $\tilde{J_k} = J - i$. This condition is precisely ensured by rejecting any proposed extensions $I^\ast_{k+1}$ for which $\mathbb{1}_{\text{sub-U-turn}}(I_{k+1}^\ast, \theta, \rho) = 1$. Remarkably, this condition is also sufficient to show the needed symmetry property.                       
\end{remark}

\begin{remark}[Computational Cost of Orbit Selection] \label{rmk:orbit-selection-complexity} In this remark, we discuss the computational cost of the random doubling procedure used by NUTS to generate $J$.   The left-most index at the $k$-th step of the procedure is $B_{k} = \sum_{l=1}^{k} 2^{l-1} \mathds{1}_{\{ I_{l}^* = J_l - |J_l|\}}$. By induction, one can prove that \[
J_{k} = \{-B_{k}, \dots, 2^{k-1} -1-B_{k} \} = [1:2^{k-1}] - (B_{k}+1) \;.
\] Given that $J_k$ has no sub-U-turns, we need to ensure that $J_{k+1}$ satisfies this property as well. In the above notation, the possible sets of intermediate leapfrog iterates corresponding to $J_{k+1}$ correspond exactly to the sets 
\begin{equation}
\{\Phi_h^i(\theta, \rho)\}_{i \in [m\cdot 2^l : (m+1)2^l -1] - (B_{k+1}+1) } , \quad l \in \{ 1, \dots, k \}, \, m \in \{ 0, 1, \dots, 2^{k+1-l} - 1 \} \label{eq:intermediate-subset-enmeration}
\end{equation} where as before ${[m : n ] = \{m, m+1, \dots, n-1, n \}}$. 

Since $J_k$ has no sub-U-turns by induction, when generating an orbit we need only check that the intermediate sets corresponding to subsets of the proposed extension $I_{k+1}^\ast$ exhibit no U-turns. Moreover, since the U-turn condition involves only the endpoints of the set of leapfrog iterates, the cost of checking whether or not a given set of leapfrog iterates satisfies the U-turn condition is linear in the size of the set (i.e., $\mathcal{O}(n)$ for leapfrog paths of size $n$). 

Therefore, the total cost of checking whether $I_{k+1}^*$ satisfies the sub-U-turn condition is upper bounded by the number of sets we need to check - which is upper bounded by the number of sets enumerated in (\ref{eq:intermediate-subset-enmeration}). Since each choice of $l$ above corresponds to $2^{l-1}$ such sets the total cost for checking that $I_{k+1}^*$ satisfies the sub-U-turn property is therefore upper bounded by $\sum_{l=1}^{k} 2^{l-1} = 2^{k}-1$. Since returning a set $J$ of size $|J| = M = 2^k$ requires checking the sub-U-turn property for $I_1^{\star}, \dots, I_{k+1}^*$, the total computational cost of producing a set of this size is $O(M)$. 

Compare this to the computational cost of the simpler strategy of extending $J_k$ one point at a time by sampling $I_{k+1}^*$ uniformly from $\{\min(J_k) -1, \max(J_k)+1 \}$. For such a strategy, the cost of appending $I_{k+1}^*$ is again of order $|J_k|$ and therefore to return a set $J$ with $|J| = M$ has computational cost $O(M^2)$. Reducing the cost of producing large orbits motivates the random doubling procedure used in NUTS. For more details on the computational cost of the random doubling procedure used in NUTS as well as other aspects of the implementation, see Appendix A in \cite{betancourt2017conceptual}.

\begin{comment}
We can consider $I_{k+1}$ in the same way. That is, to every point $i \in I_{k+1}$ also corresponds a hypothetical sequence of binary choices $U_1'', \dots, U_k''$ which represents a sequence of intermediate sets $I'_1, \dots, I'_k$ which would have produced from $\Phi_h^i(\theta, \rho)$ the same set of leapfrog iterates $L((\theta, \rho), I_{k+1})$ as are represented by $I_{k+1}$ starting at $(\theta, \rho)$. In order to preserve the property $P_h(J \mid \theta, \rho) = P_h(J-i \mid \Phi_h^i(\theta, \rho))$ we need only check that none of these hypothetical intermediate sets would have triggered the No-U-Turn condition. If $I_{k+1}$ satisfies this property, we set $J_{k+1} = J_k \cup I_{k+1}$. Otherwise, we return $J_{\theta, \rho} = J_k$. We continue this procedure until $J_k$ satisfies the U-Turn condition, we encounter a rejected proposal, or $J_k$ reaches some user-specified maximum size. As a consequence of this careful construction, the orbit selection kernel satisfies $P_h(J \mid \theta, \rho) = P_h(J-i \mid \Phi_h^i(\theta, \rho))$.
\end{comment}
\end{remark}

\subsection{The Apogee-to-Apogee Path sampler as a GIST sampler}\label{sec:Apogee-to-Apogee}

Here we show that the Apogee-to-Apogee Path Sampler (AAPS) introduced in \cite{SherlockUrbasLudkin2023Apogee} is a GIST sampler.  The idea of AAPS is to partition the discrete path  $\{ \Phi_h^i (\theta, \rho) \}_{i \in \mathbb{Z}}$ originating from the current state $(\theta, \rho) \in \mathbb{R}^{2d}$ into segments $\{S_i(\theta, \rho) \}_{i \in \mathbb{Z}}$. In particular, the $i$-th segment $S_i(\theta, \rho)$ consists of points along the discrete path  that lie between consecutive local maxima (aka ``apogees'') of the underlying  potential energy $U(\theta)$.   The number of leapfrog steps is then chosen based on this partition of discrete paths.   

In order to write AAPS as a GIST sampler, we first explain AAPS more precisely starting with the partitioning mentioned above. Proceeding both forwards and backwards in time from $(\theta, \rho)$  via leapfrog steps, consecutive outputs belong to the same segment if the potential $U$ is either instantaneously increasing at both points or instantaneously decreasing at both points. Otherwise, an apogee occurs between this pair of points, and hence, a new segment is defined. Iterating this procedure produces a two-sided sequence of segments: \[
\dots, S_{-1}(\theta, \rho), S_0(\theta, \rho), S_1(\theta, \rho), \dots 
\] 
where $ S_0(\theta, \rho)$ is the initial segment.  Following the notation of \cite{SherlockUrbasLudkin2023Apogee}, let $S_\#((\theta, \rho), (\theta', \rho'))$ be the index of the segment started at $(\theta, \rho)$ containing the leapfrog iterate $(\theta', \rho')$. For instance, $S_\#((\theta, \rho), (\theta, \rho)) = 0$ since the starting point is always in the initial segment $S_0(\theta, \rho)$ and $S_\#((\theta, \rho), (\theta', \rho')) = k$ for every $(\theta', \rho') \in S_k(\theta, \rho)$ and $k \in \mathbb{Z}$. Additionally, let $S_{a:b}(\theta, \rho) = \bigcup_{i = a, \dots, b} S_i(\theta, \rho)$ for $a, b \in \mathbb{Z}$ such that $a \le b$.

A positive integer $K$ and a user-defined weight function $w: \mathbb{R}^{4d} \to [0, \infty)$ are required to fully instantiate AAPS.  The latter assigns weights to points within selected sets of segments. From the current state $(\theta, \rho)$, AAPS chooses among the $K+1$ sets of segments, $S_{0:K}, S_{-1:K-1}, \dots, S_{-K:0}$, uniformly at random. These sets of segments are all possible unions of $K+1$ consecutive segments whose union contains the current state. Having selected such a set of segments $S_{-c:K-c}$, AAPS then randomly selects a leapfrog iterate $(\theta', \rho') \in S_{-c:K-c}$ as a proposal with weight $w((\theta, \rho), (\theta', \rho'))$. Finally, this proposal is Metropolized with Metropolis-Hastings acceptance probability
\begin{equation} \label{eq:acceptanceprobability_AAPS}   
1 \wedge \left( e^{-\Delta H(\theta,\ \ \rho)} \,  \frac{ w((\theta', \rho'), (\theta, \rho)) \ \sum_{(\Tilde{\theta}, \, \Tilde{\rho}) \in S_{-c:K-c}} w((\theta, \rho), (\Tilde{\theta}, \Tilde{\rho}))}{ w((\theta, \rho), (\theta',\rho')) \ \sum_{(\Tilde{\theta}, \, \Tilde{\rho}) \in S_{-c:K-c}} w((\theta', \rho'), (\Tilde{\theta}, \Tilde{\rho})) } \right) \;. \end{equation}
The weight function can be chosen to preferentially select proposals according to various desiderata. For instance, one might take $w((\theta,\rho),\,(\theta',\rho')) = e^{-H(\theta',\ \rho')}$ to select points which will always be accepted under the Metropolis-Hastings step, or $w((\theta, \rho), (\theta',\rho')) = e^{-H(\theta',\ \rho')} || \theta - \theta'||^2$ to bias toward proposals which are farther from the current state. In addition to these, several other choices of weight functions are suggested in \cite{SherlockUrbasLudkin2023Apogee}. 

We are now in position to write AAPS as a GIST sampler. Let $\mathbb{A} = \mathbb{Z}^2$ with background measure given by the counting measure, and write the algorithm tuning parameter as a pair $\alpha = (c, i) \in \mathbb{A}$. The component $c$ represents the choice of set of segments $S_{-c:K-c}$ while the component $i$ represents the choice of leapfrog iterate $\Phi_h^i(\theta, \rho)$ within this set of segments.  Define 
\begin{equation} \label{eq:p_aaps}
p(c,i \mid \theta, \rho) 
= \frac{1}{K+1} 
\ \frac{w((\theta, \rho), \Phi_h^i(\theta, \rho)) 
        \ \mathds{1}_{S_{-c:K-c}(\theta,\, \rho)}(\Phi_h^i(\theta, \rho))}
        {\sum_{j \in \mathbb{Z}} w((\theta, \rho), \Phi_h^j(\theta, \rho)) 
         \ \mathds{1}_{S_{-c:K-c}(\theta,\, \rho)}(\Phi_h^j(\theta, \rho))} \;,
\end{equation} 
where $c \in [0:K]$ and $i \in [-c:K-c]$.  Define the measure-preserving involution $G$ in \eqref{eq:G} by 
\begin{equation} \label{eq:G_aaps}
G : (\theta, \rho, c, i) 
\mapsto  
\Big(\mathcal{S} \circ \Phi_h^i(\theta, \rho),
\ c + S_{\#}\big((\theta, \rho), \, \mathcal{S} \circ \Phi_h^i(\theta, \rho)\big), 
\ i \Big) \;.
\end{equation} 
With these specifications, the corresponding acceptance probability in \eqref{eq:acceptanceprobability} simplifies to \eqref{eq:acceptanceprobability_AAPS} and the corresponding GIST sampler in Algorithm~\ref{algo:general-self-tuning-step} reduces to AAPS.   Once one confirms that the map $G$ in \eqref{eq:G_aaps} is a measure-preserving involution, the following corollary of Theorem~\ref{thm:gist_reversibility} holds.

\begin{corollary} \label{cor:AAPS-reversible}
The transition kernel of the Apogee-to-Apogee Path Sampler is reversible with respect to $\mu$.
\end{corollary}

A proof of Corollary~\ref{cor:AAPS-reversible} is provided in Appendix~\ref{app:proofs}.

\section{Truncation of an infinite-dimensional Gaussian measure} \label{sec:nealsexample}

This section illustrates that the GIST samplers described in Section~\ref{ex:exact_stHMC} can achieve a similar mean-squared jump distance (MSJD) as well-tuned randomized HMC from Section~\ref{ex:exact_rHMC} on a target distribution that can be interpreted as a truncation of an infinite-dimensional Gaussian measure \cite{BePiSaSt2011,BoEb2020}; see Remark~\ref{rmk:applications} for applications of this class of target measures.  This is a worst-case example, because the corresponding Hamiltonian dynamics is highly oscillatory \cite{petzold_jay_yen_1997}. More precisely, the target distribution is a $d$-dimensional centered Normal distribution with covariance matrix given by 
\begin{equation}\label{eq:sigma_diag}
\Sigma = \operatorname{diag}(\sigma_1^2, \dots, \sigma_d^2) \;, 
\quad \text{where} \quad  \sigma_i 
= \frac{i}{d} \quad \text{for $i \in \{1, \dots, d \}$} \;.
\end{equation}
This example has traditionally been used to illustrate the importance of path-length randomization to avoid slow mixing due to periodicities or near-periodicities in the  underlying Hamiltonian dynamics \cite{Ne2011,BoSa2017,BoEb2022}.  Here we use this worst-case example to demonstrate the efficacy of GIST samplers based on the U-turn conditions defined by \eqref{eq:ct_angle} and \eqref{eq:ct_dist}.

\begin{figure}[t]
    \centering
    \includegraphics[width=0.48\textwidth]{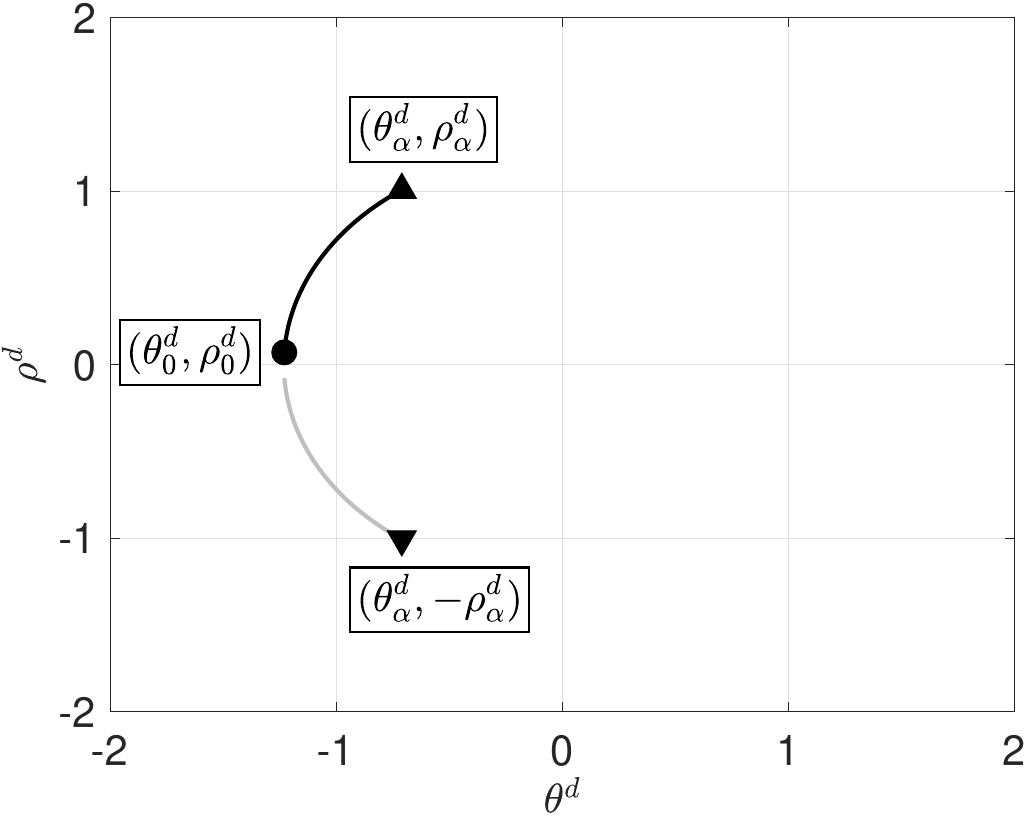} ~~~
    \includegraphics[width=0.48\textwidth]{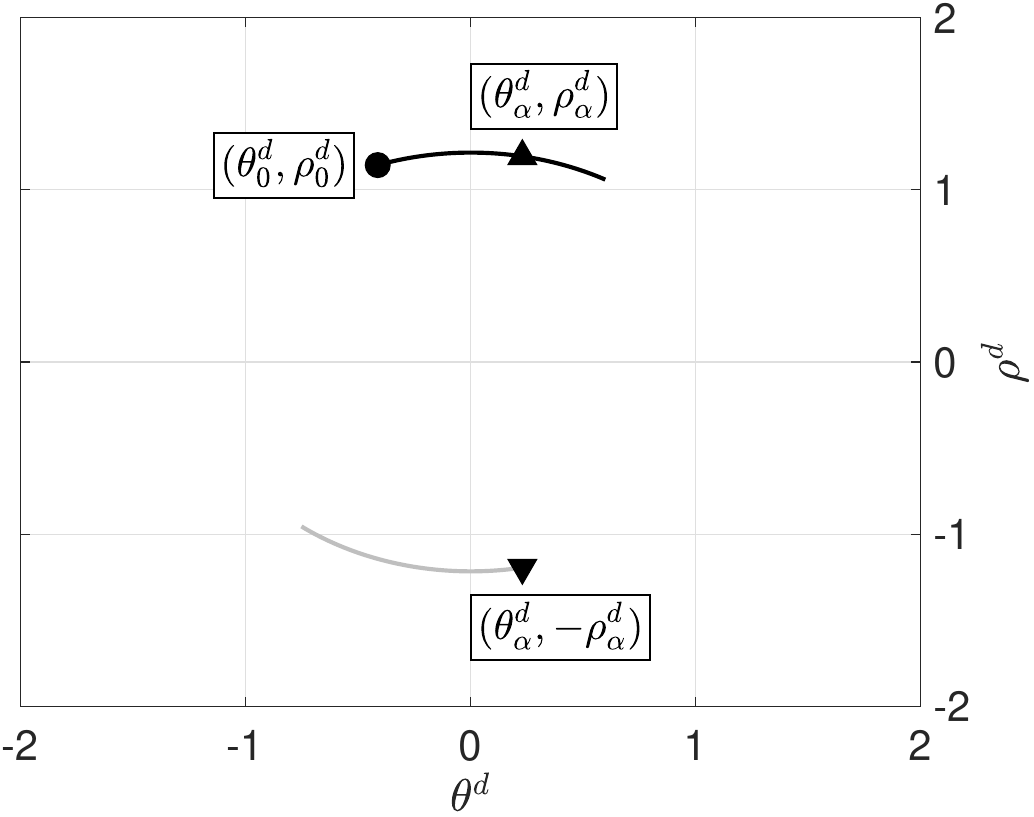} 
    \caption{\it {\bfseries GIST Sampler based on Exact Hamiltonian Flow.} This figure plots forward and backward trajectories for evaluating \eqref{eq:ct_angle} in the phase space of the least constrained coordinate of the $d$-dimensional Gaussian measure with covariance matrix given in \eqref{eq:sigma_diag} and $d=10^3$.    The left panel illustrates a rejected proposal (due to $\alpha>\tau_2$) while the right panel illustrates an accepted proposal.}
    \label{fig:stHMC_illustration}
\end{figure}

For this model, the exact solution to the corresponding Hamiltonian dynamics $(\theta_t, \rho_t)$ at time $t \ge 0$ from initial condition $(\theta, \rho)$ is given in components by 
\[
\theta_t^i = \cos\left( \frac{t}{\sigma_i} \right) \theta^i + \sigma_i \sin\left( \frac{t}{\sigma_i} \right) \rho^i 
\quad \textrm{and} \quad
\rho_t^i = - \frac{1}{\sigma_i} \sin\left( \frac{t}{\sigma_i} \right) \theta^i + \cos\left( \frac{t}{\sigma_i} \right) \rho^i,  
\] 
for $i \in \{1, \dots, d \}$.  Using these solutions, the evaluation of the U-turn path lengths in \eqref{eq:ct_angle} or \eqref{eq:ct_dist} reduces to finding the first positive root of a scalar function of time, which itself crucially relies on a good initialization.  

To be precise, we provide a complete description of the initialization procedure for the case of the U-turn path length given in \eqref{eq:ct_angle}; a similar initialization is used for \eqref{eq:ct_dist} and therefore omitted.  Our goal is to find the first positive root of $f$, i.e., 
\[
\tau = \inf\{ t \ge 0 ~:~ f(t) = 0 \} \;.
\] 
where we have defined the function $f(t)$ of time by
\[
f(t) = \sum_{i=1}^d f_i(t) \;,
\quad 
\text{where} \quad
f_i(t) = -\frac{1}{\sigma_i} \sin\left( \frac{t}{\sigma_i} \right)  \theta^i \rho^i + \cos\left( \frac{t}{\sigma_i} \right) (\rho^i)^2 \;. 
\]
The first positive root of $f_i$ has the analytic form
\[
 \tau_i = \arctan\left( \frac{\rho^i_0}{\theta^i_0} \right)  + k^{\star} \pi \;,  
\quad 
\text{where} \quad
k^{\star} = \min\!\left\{ k \in \mathbb{N} ~:~ \arctan\left( \frac{\rho^i_0}{\theta^i_0} \right)  + k \pi \ge 0  \right\} \;. 
\]
The mean of $\{ \tau_i \}_{i=1}^d$ is used to seed the root solver for finding $\tau$.     
Using the above root finding procedure to evaluate $\tau$ per transition step, the numerically estimated mean-acceptance probability (mean $a_{\mathrm{GIST}}$),  mean-squared jump distance (MSJD), and mean path length (mean $\tau$) for $d=1000$ and using $10^5$ transition steps is summarized in Table~\ref{tbl:exact-results}.
\begin{table}
\begin{center}
\begin{tabular}{c|c|c|c}
& mean $a_{\mathrm{GIST}}$ & MSJD & mean $\tau$ \\
\hline
randomized HMC & 100\% & 429.81 & 1.00 \\
exact GIST sampler using angle U-turn condition \eqref{eq:ct_angle} & 97.4\% & 174.85 &  0.44 \\
exact GIST sampler using distance U-turn condition  \eqref{eq:ct_dist} & 94.4\%  & 573.15 &  1.16
\end{tabular}
\end{center}
\caption{\textit{{\bfseries Comparison of exact GIST samplers}.}}\label{tbl:exact-results}
\end{table}
Randomized HMC is operated using the optimal choice of mean path length for maximizing expected square jump distance in this example, which corresponds to the standard deviation of the least constrained coordinate \cite[Section 4]{BoSa2017}.  The self-tuned HMC algorithm based on \eqref{eq:ct_dist} slightly outperformed exact randomized HMC in terms of MSJD, while the one based on \eqref{eq:ct_angle} performed worse.  Remarkably, neither degenerated in high dimension.  This high-dimensional example demonstrates both the efficacy of the U-turn conditions and the leniency of the corresponding acceptance probability in the self-tuned HMC algorithm.

\begin{remark} \label{rmk:applications}
These numerical findings are of independent interest, since they motivate using GIST samplers based on U-turn avoiding conditions to sample from perturbed Gaussian measures on Hilbert spaces.   This class of target measures is relevant to several important applications including Path Integral Molecular Dynamics \cite{ChandlerWolynes,
Miller2005a,
  Habershon2013,lu2020continuum}, Transition Path
Sampling  \cite{ReVa2005,pinski2010transition,bolhuis2002,Miller2007}, and Bayesian statistical inverse problems for Hilbert spaces \cite{kaipio2005statistical,stuart2010inverse,dashti2017bayesian,borggaard2020bayesian}. 
Since the corresponding Hamiltonian dynamics is potentially highly oscillatory in high modes \cite{petzold_jay_yen_1997}, in actual numerical implementations, preconditioning \cite{BoSaActaN2018} or strongly stable integrators \cite{Korol2020} are necessary to be able to choose the step size
independently of the dimension.  
\end{remark}

\newcommand{\pos}[2]{#1^{(#2)}}
\begin{algorithm}[t]
\small
\begin{flushleft}
$\textbf{GIST}(\theta, h, \Sigma, \mu, p)$
\vspace*{2pt}
\hrule
\vspace*{2pt}
\begin{tabular}{ll}
$\theta \in \mathbb{R}^d$ & initial position
\\[2pt]
$h \in (0, \infty)$ & step size
\\[2pt]
$\Sigma \in \mathbb{R}^{d \times d}$ & symmetric, positive definite mass matrix
\\[2pt]
$\mu(\theta)$ & target density (log density evaluation \& gradient)
\\[2pt]
$p(L \mid \theta, \rho)$ & conditional steps distribution (sampler \& log density evaluation)
\end{tabular}
\vspace*{4pt}
\hrule
\vspace*{6pt}
$\pos{\theta}{0} = \theta$  \hfill (initialize)
\\[10pt]
{\footnotesize (GIBBS)} \\[2pt]
$\pos{\rho}{0} \sim \textrm{normal}(0, \Sigma)$ \hfill (Gibbs sample momentum)
\\[4pt]
$L \sim p(\cdot \mid \pos{\theta}{0}\!,\, \pos{\rho}{0})$ \hfill (Gibbs sample number of steps)
\\[10pt]
{\footnotesize (METROPOLIS)} \\[2pt]
for $\ell$ from $0$ to $L - 1$ (inclusive):  \hfill ($L$ leapfrog steps)
\\[-6pt]
\null \qquad $\pos{\rho}{\ell + 1/2} = \pos{\rho}{\ell} + \frac{h}{2} \cdot \nabla \log \mu(\pos{\theta}{\ell})$ \hfill (half step momentum)
\\[-6pt]
\null \qquad $\pos{\theta}{\ell + 1} = \pos{\theta}{\ell} + h \cdot \Sigma^{-1} \cdot \pos{\rho}{\ell + 1/2}$ \hfill (full step position)
\\[-6pt]
\null \qquad $\pos{\rho}{\ell + 1} = \pos{\rho}{\ell + 1/2} + \frac{h}{2} \cdot \nabla \log \mu(\pos{\theta}{\ell + 1})$ \hfill (half step momentum)
\\[6pt]
($\theta', \rho', L') = (\pos{\theta}{L}, -\pos{\rho}{L}, L)$  \hfill (proposal flips momentum)
\\[4pt]
$u \sim \textrm{uniform}([0, 1])$ \hfill (sample acceptance probability)
\\[4pt]
if
$u < \frac{\displaystyle \widehat{\mu}(\theta'\!\!,\, \rho'\!\!,\,  L' )}
         {\displaystyle \widehat{\mu}(\pos{\theta}{0}\!,\, \pos{\rho}{0}\!,\, L )}$
\hfill (Metropolis accept condition)
\\
\null \quad return $  \theta'$ \hfill (accept)
\\[4pt]
else
\\[-12pt]
\null \quad return $\null \theta$ \hfill (reject) \\[4pt]
\vspace*{2pt}
\hrule
\caption{\it {\bfseries GIST sampler for path-length self-tuning}.  The GIST sampler for path-length self-tuning differs from standard HMC in sampling the number of steps each iteration and then adjusting the acceptance probability to ensure detailed balance.  Note, the Gibbs steps for refreshment of both momentum and number of steps are exact draws from the corresponding conditional distributions.}
\label{algo:self-tuning-hmc}
\end{flushleft}
\end{algorithm}

\section{Path length sampling to avoid U-turns} \label{sec:step-distro} \label{sec:u-turn-avoiding}

We now turn to an extended example of a novel GIST sampler that incorporates a U-turn avoiding condition. We focus on locally adapting the number of leapfrog steps $L$, and fix the step size $h$ and mass matrix $\Sigma$ throughout. For each HMC step, the GIST sampler generates $L$ probabilistically according to a tuning parameter distribution $p(L \mid \theta, \rho)$. Several such distributions are described in this section, and then empirically evaluated in Section~\ref{sec:experiments}.

Algorithm~\ref{algo:self-tuning-hmc} provides pseudocode for a general GIST sampler adapting the number of leapfrog steps.  Note that the GIST acceptance ratio in Algorithm~\ref{algo:self-tuning-hmc}  can be factored as in Algorithm ~\ref{algo:general-self-tuning-step}
\[
\frac{\displaystyle \widehat{\mu}\!\left(\theta', \rho',  L' \right)}
         {\displaystyle  \widehat{\mu}\!\left(\pos{\theta}{0}, \pos{\rho}{0}, L \right)}
\ = \ 
\frac{\displaystyle \mu\!\left(\theta'\right)}
     {\displaystyle \mu\!\left(\pos{\theta}{0}\right)}
\cdot 
\frac{\displaystyle p_\rho\!\left(\rho' \right)}
     {\displaystyle p_\rho\!\left(\pos{\rho}{0} \right)}
\cdot
\frac{\displaystyle p\!\left(L' \mid \theta', \rho' \right)}
     {\displaystyle p\!\left(L \mid \pos{\theta}{0}, \pos{\rho}{0}\right)} 
= e^{-\Delta H(\pos{\theta}{0},\ \pos{\rho}{0})} \frac{p(L' \mid \theta', \rho')}{p(L \mid \theta^{(0)}, \rho^{(0)})} \, ,
\]
where we used the shorthand $p_\rho(\rho) = \textrm{normal}(\rho \mid 0, \Sigma)$ to denote the momentum probability density function, and $p(L \mid \theta, \rho)$ to denote the conditional path length probability mass function.

\subsection{Step distributions avoiding U-turns} 

\begin{figure}[t!]
\usetikzlibrary{arrows.meta, angles, quotes, calc}
\begin{center}
\begin{tikzpicture}[>=Stealth]
  % Define points along the arc
  \coordinate (A) at (0,0);
  \coordinate (B) at (2,2);
  \coordinate (C) at (4,1);
  \coordinate (D) at (6,3); % Second to last point
  \coordinate (E) at (8,2); % Last point

  \node at (A) [below=0.125cm] {$\pos{\theta}{0}$};
  \node at (B) [above=0cm] {$\pos{\theta}{1}$};
  \node at (C) [below=0.125cm] {$\pos{\theta}{2}$};
  \node at (D) [above=0cm] {$\pos{\theta}{3}$};
  \node at (E) [below=0.125cm] {$\theta^*$};

  % Draw trajectory with arrows
  \draw[->] (A) -- (B);
  \draw[->] (B) -- (C);
  \draw[->] (C) -- (D);
  \draw[->,style=dashed] (D) -- (E);

  % Place solid circles at discretized positions
  \foreach \point in {A,B,C,D,E}
    \fill (\point) circle (2pt);

  % Dotted line from first to second to last point
  \draw[dotted] (A) -- (D);

  % Auxiliary point for angle calculation
  % This creates an invisible point extending the dotted line beyond 'D' for angle drawing
  \coordinate (F) at ($(A)!-1.2!(D)$); % Extend line beyond D for angle marking

  % Arc for angle indication
  \pic[draw, ->, "$\omega$", angle eccentricity=1.5, angle radius=1cm] {angle = F--D--E};
\end{tikzpicture}
\end{center}
\caption{\it {\bfseries U-turn condition.}  A Hamiltonian trajectory of positions (not momenta) in two dimensions, consisting of three leapfrog steps plus a potential fourth step $\theta^*$.  The dotted line connects the initial position $\pos{\theta}{0}$ to the current position $\pos{\theta}{3}$.  The dashed line connects the current position to the next potential position $\theta^*$ and approximately runs in the direction of the current momentum $\pos{\rho}{3}$.  The trajectory is extended one step to $\theta^*$ if the next step moves away from the initial position, which requires the absolute value of the angle $\omega$ between the dotted line and the dashed line to be greater than $\pi/2$, which arises when $(\pos{\theta}{3} - \pos{\theta}{0}) \cdot (\theta^* - \pos{\theta}{3}) > 0$, or approximately, when $(\pos{\theta}{3} - \pos{\theta}{0}) \cdot \pos{\rho}{3} > 0.$}
    \label{fig:u-turn-condition}
\end{figure}

To make our sampler concrete, a specific distribution over the number of steps must be defined.  We evaluate a few related choices, all of which are motivated by the main idea underlying NUTS, namely that it is wasteful to continue extending the Hamiltonian trajectory once it is moving back toward the initial point \cite{HoGe2014}. Figure~\ref{fig:u-turn-condition} illustrates the U-turn condition, which is made precise in Equation~(\ref{eq:uturn-cond}).  

Let $\mathrm{U}(\theta, \rho)$ be the maximum number of leapfrog steps starting from $(\theta, \rho)$ that can be taken before a U-turn in the sense of \eqref{eq:ct_dist}. That is, 
\begin{equation}\label{eq:uturn-cond}
\mathrm{U}\!\left( \theta\!,\, \rho \right)
= \textrm{arg min}_{n \in \mathbb{N}} \ \left(\pos{\theta}{n} - \pos{\theta}{0}\right)^\top \cdot \pos{\rho}{n} < 0,
\end{equation}
where $(\pos{\theta}{n}\!, \pos{\rho}{n}) = \Phi_h^n(\theta, \rho)$. 

\begin{figure}[t!]
    \centering
    \null\vspace{12pt}
\begin{tikzpicture}[node distance=1.5cm and 0.75cm]
    % Top row nodes
    \node[circle,draw,minimum size=1cm] (LM) {\footnotesize $L{-}N$};
    \node[minimum size=1cm,right of=LM] (dots1) {$\cdots$};
    \node[circle,draw,minimum size=1cm,right of=dots1] (0) {$0$};
    \node[minimum size=1cm,right of=0] (dots2) {$\cdots$};
    \node[circle,draw,minimum size=1cm,right of=dots2] (L) {$L$};
    \node[minimum size=1cm,right of=L] (dots3) {$\cdots$};
    \node[circle,draw,minimum size=1cm,right of=dots3] (N) {$M$};
    
    % 2nd row nodes
    \node[below = 0.05cm of LM] (n_a1) {$\strut$};
    \node[below = 0.05cm of 0] (theta_rho) {$\strut (\theta, \rho)$};
    \node[below = 0.05cm of L] (theta_rho_star) {$\strut (\theta', \rho')$};
    \node[below = 0.05cm of N] (n_a2) {$\strut$};

    % 3rd row arrow
    \node[below = -0.25cm of n_a1] (n_a3) {$\strut$};
    \node[below = -0.25cm of theta_rho] (start_arrow2) {$\strut$};
    \node[below = -0.25cm of theta_rho_star] (n_a4) {$\strut$};
    \node[below = -0.25cm  of n_a2] (end_arrow2) {$\strut$};
    \draw[->, line width=1pt] (start_arrow2) -- (end_arrow2) node[midway, below] {$M = \mathrm{U}(\theta, \rho)$};
    
    % 4th row arrow
    \node[below = 0.05cm of n_a4] (start_arrow1) {$\strut$};
    \node[below = 0.05cm of n_a3] (end_arrow1) {$\strut$};
    \draw[->, line width=1pt] (start_arrow1) -- (end_arrow1) node[midway, below] {$N = \mathrm{U}(\theta', \rho')$};
\end{tikzpicture}
    \caption{\it {\bfseries Self-tuned steps proposal.} Starting from the initial position and momentum $(\theta, \rho)$, the algorithm makes $M$ forward leapfrog steps until a U-turn.  It then Gibbs samples a number of steps $L$ between 0 and $M$ for which the leapfrog integrator produces the proposal $(\theta', \rho')$.  Then $N$ backward leapfrog steps are taken until a U-turn.
   If $L - N > 0,$ the proposal is rejected.  Computationally, the algorithm takes $M + N - L$ unique leapfrog steps.}
    \label{fig:adaptive-u-turns}
\end{figure}  
Figure~\ref{fig:adaptive-u-turns} shows a single step of the algorithm.  As illustrated, the number of steps $L$ is sampled from a conditional distribution of steps that incorporates the number of steps to a U-turn, $\mathrm{U}(\theta, \rho)$.  %To preserve detailed balance, the trajectory from the selected point backward in time is evaluated until it makes a U-turn and the probability of selecting the initial state (i.e., the same number of steps) is used to balance the selection.  To avoid rejection, this requires sampling zero or more steps backward in time before the initial position.

\subsection{Conditional distribution of steps} 

Here we consider a few closely related tuning parameter distributions for the number of steps $L$.  

\subsubsection{Steps generated uniformly}

The most obvious choice for a conditional distribution over the number of steps is uniform between 1 and $\mathrm{U}(\theta, \rho)$,
\begin{equation} \label{eq:uniform_tuning_distribution}
p(L \mid \theta, \rho) \ = \ \textrm{uniform}(L \mid[1:\mathrm{U}(\theta, 
\rho)]) 
 \ = \  \frac{1}{\mathrm{U}(\theta, \rho)} \mathbb{1}_{[1:\mathrm{U}(\theta, \rho)]}(L)
\,.
\end{equation}
%where the bounds are read inclusively.

\subsubsection{Steps generated uniformly from later states}

One of the strategies that the revised version of NUTS \cite{betancourt2017conceptual} and the Apogee-to-Apogee Path Sampler \cite{SherlockUrbasLudkin2023Apogee} use to take longer jumps is to bias the selection of a proposal toward the end of the Hamiltonian trajectory.  Neal \cite{neal1994improved} developed an HMC method that encouraged longer jumps by restricting proposals to the later states in a trajectory and taking a multinomial approach; this can also be shown to be an instance of GIST.  We use a no-U-turn condition rather than a fixed size, then follow Neal in restricting the uniform distribution in \eqref{eq:uniform_tuning_distribution} to the latter part of the trajectory by changing the lower bound from 1 to something greater: 
\begin{align}\label{eq:biased_selection_uniform_tuning_distribution}
p(L \mid \theta, \rho)
\ = \  \textrm{uniform}(L \mid [ \textrm{max}(1, \, \lfloor \psi \mathrm{U}(\theta, \rho)\rfloor):\mathrm{U}(\theta, \rho)]) \, ,
%\&= \frac{1}{\lfloor (1 - \psi) U(\theta, \rho)\rfloor} \ \mathbb{1}_{\big[\textrm{max}(1, \, \lfloor \psi U(\theta, \rho)\rfloor), U(\theta, \rho)\big]}(L) 
\end{align}
where $\psi \in (0,1]$.  The choice of $\psi=0$ corresponds to the uniform distribution of the previous section.  
%We will also evaluate lower bounds of $\lfloor \frac{1}{2} \cdot M \rfloor$ and $\lfloor \frac{3}{4} \cdot M \rfloor$.  
With smaller $\psi$, the proposed trajectory lengths will be longer in expectation, but the acceptance probability within GIST might be lower.

\subsubsection{Binomial step generation}

Non-uniform distributions may also be used.  For example, a binomial distribution for the number of steps could be defined for a fixed $\chi \in (0, 1)$ as 
\begin{equation}
    p(L \mid \theta, \rho)
    = \binom{\mathrm{U}(\theta, \rho)}{L} \chi^L (1- \chi)^{\mathrm{U}(\theta, \rho) - L} 
    = \textrm{binomial}(L \mid \mathrm{U}(\theta, \rho), \chi) \, .
\end{equation}
In practice, the binomial method led to low acceptance rates, so the results are not reported in Section~\ref{sec:experiments}.

\subsection{Empirical evaluation}\label{sec:experiments}

\subsubsection{Models evaluated}

The test models considered are described in Appendix~\ref{sec:test_models}. Apart from the normal and Rosenbrock distributions, they are all drawn from the \texttt{posteriordb} package \cite{posteriordb2023}.  Short for posterior database, \texttt{posteriordb} contains a wide range of Bayesian models applied to real data sets.

\begin{figure}[t]
    \centering
    \null \hfill
\includegraphics[width=0.45\textwidth]{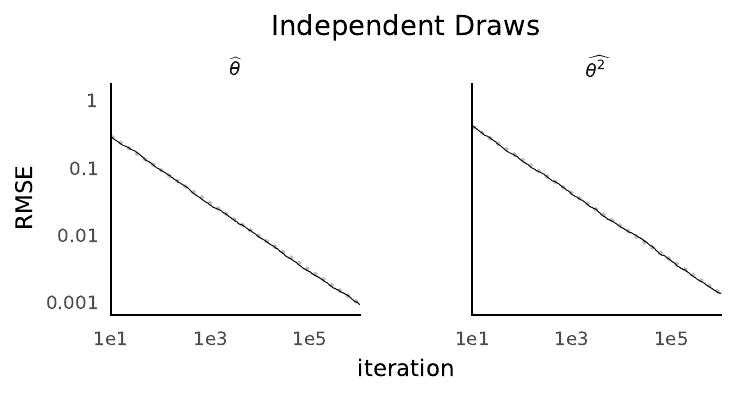}
\hfill
\includegraphics[width=0.45\textwidth]{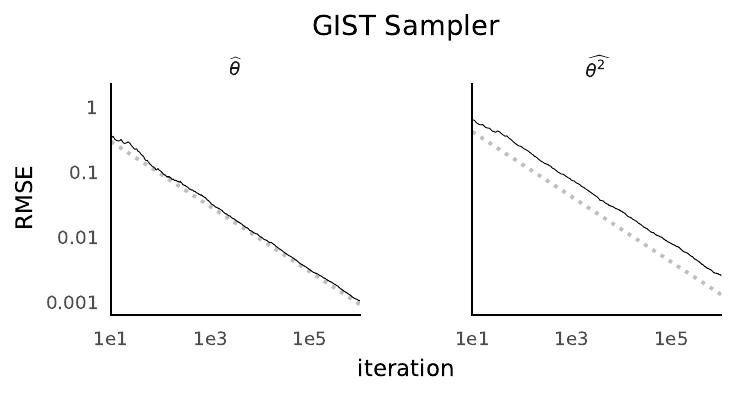}
\hfill \null
\vspace*{-9pt}
\caption{\it {\bfseries Learning curve validation.} The root mean square error (RMSE) in estimating the first and second moments ($\theta, \theta^2$) for a 100-dimensional standard normal initialized randomly from the target, with i.i.d.~draws (left) and the GIST sampler based on \eqref{eq:uniform_tuning_distribution} with a step size of 0.25 (right).  The dotted line shows the RMSE expected for independent draws.}
\label{fig:learning-curve}
\end{figure}

\subsubsection{Learning curve}

Figure~\ref{fig:learning-curve} plots expected absolute error in estimates for the first and second moment against iteration for the GIST sampler based on \eqref{eq:uniform_tuning_distribution} for a 100-dimensional standard normal model as a simple validation that such samplers target the correct distribution. The plot shows that error decreases as expected and the efficiency is comparable to Monte Carlo estimates obtained with i.i.d.\ draws.

\subsubsection{Effect of step size and path fraction}

\begin{figure}[t]
\begin{center}
\includegraphics[width=0.8\textwidth]{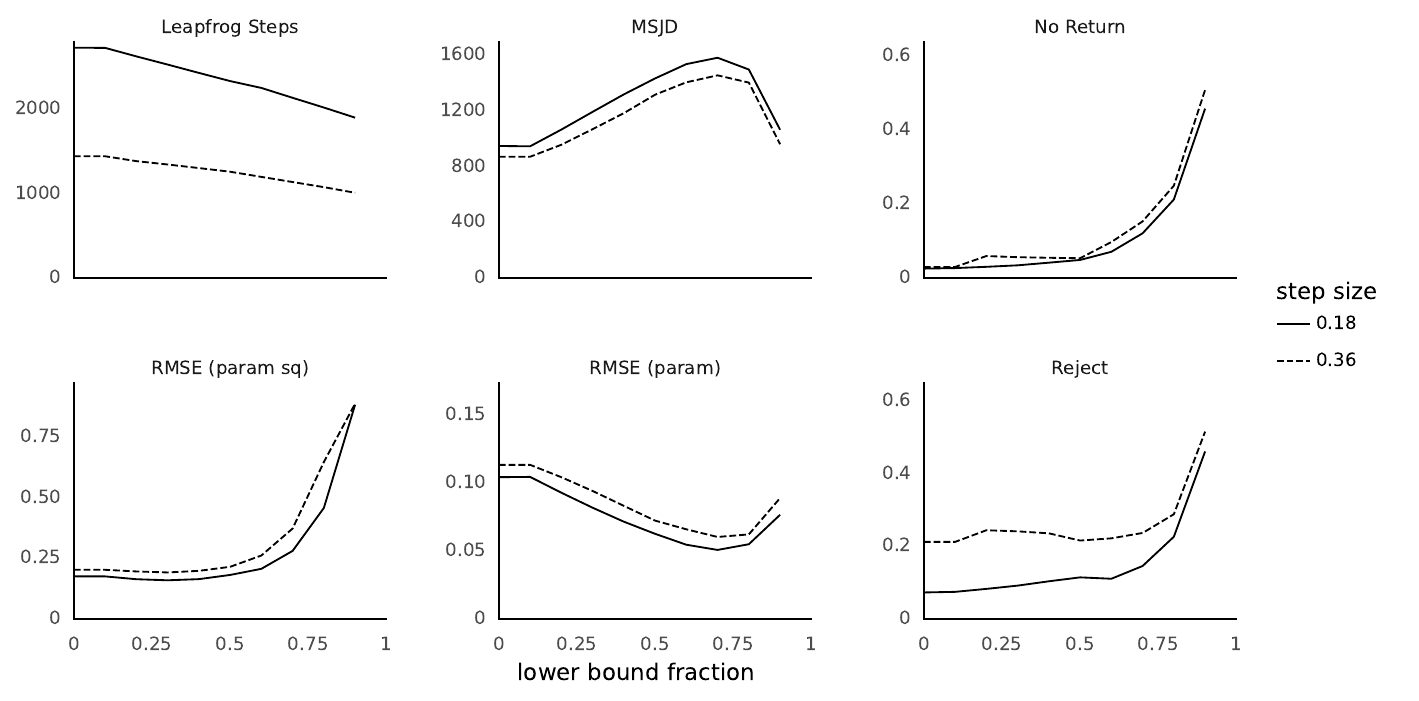}
\end{center}
\caption{\it {\bfseries Performance vs.\ step size and lower bound fraction}.  The plots show averages (over 500 repetitions) of sampling a 500-dimensional standard normal for 100 iterations of the GIST sampler based on \eqref{eq:biased_selection_uniform_tuning_distribution} starting from a random draw from the stationary distribution.  The $x$-axis represents the path fraction $\psi$ used in \eqref{eq:biased_selection_uniform_tuning_distribution}.  Line color indicates step sizes 0.36 (blue) and 0.18 (red).  The titles of the subplots describe the values on the $y$-axis.  The label "No Return" (top right) is for the fraction of rejections due to $L > \mathrm{U}(\theta',\rho')$, and the proportion of rejections (lower right) includes those due to $L > \mathrm{U}(\theta',\rho')$.}\label{fig:step-lb-uniform}
\end{figure}

In Figure~\ref{fig:step-lb-uniform}, the performance of the GIST sampler based on \eqref{eq:biased_selection_uniform_tuning_distribution} is shown for two step sizes, 0.36 (blue lines) and 0.18 (red lines), across a range of lower bound fractions ($x$ axis). The target in this case is the 500-dimensional standard normal distribution. The step size 0.36 is what NUTS adapted for an 80\% average Metropolis acceptance probability (Stan's default); 0.18 is the step size for roughly 95\% Metropolis acceptance.  Halving the step size roughly doubles the number of leapfrog steps taken, as shown in the upper left of the plot.  The remaining plots show that accuracy is better with a smaller step size.  

Mean square jump distance (MSJD) is also shown in Figure~\ref{fig:step-lb-uniform}; for $M$ sampling steps, it is defined  by
\begin{equation}
\textrm{MSJD}
= \frac{1}{M} \sum_{m = 1}^M \left|\pos{\theta}{m+1} - \pos{\theta}{m}\right|^2.
\end{equation}
The MSJD is an estimate of the mean squared jump distance starting from the stationary distribution, 
\begin{equation}
\textrm{ESJD}
= \mathbb{E}\!\left[\left|\pos{\theta}{1} - \pos{\theta}{0}\right|^2\right]  \qquad  \text{where $\pos{\theta}{0} \sim \mu_{\theta}$.}
\end{equation}

As $a_{\mathrm{GIST}} = 0$ if $L > \mathrm{U}(\theta', \rho')$ (see Figure~\ref{fig:adaptive-u-turns}), the rejection rate is broken down into total rejection rate and then the number of rejections due to $L > \mathrm{U}(\theta', \rho')$ --- the ``no-return" rejection rate.  Smaller step sizes more accurately preserve the Hamiltonian and thus have lower rejection rates. As the path fraction increases, the MSJD increases until the no-return rejection rate dominates rejections and it begins to decrease.

For $M$ sampling steps, the standardized root mean square error (RMSE) for an estimate $\hat{\theta}$ of parameter $\theta \in \mathbb{R}^M$ is defined by
\begin{equation}
\textrm{(standardized) RMSE} 
= 
\sqrt{\frac{1}{M} \sum_{m=1}^M
    \left| \frac{\pos{\theta}{m} - \textrm{mean}(\theta)}{\textrm{sd}(\theta)}\right|^2
},
\end{equation}
where $\textrm{mean}(\theta)$ is the mean and $\textrm{sd}(\theta)$ is the standard deviation of the parameter $\theta$. Standardizing RMSE is common to bring all the errors onto a $z$-score scale where they represent number of standard deviations away from the mean and thus make the model parameters more comparable for averaging.

\subsubsection{Evaluations for multiple models}

\begin{figure}[th]
    \centering
    \includegraphics[width=0.95\textwidth]{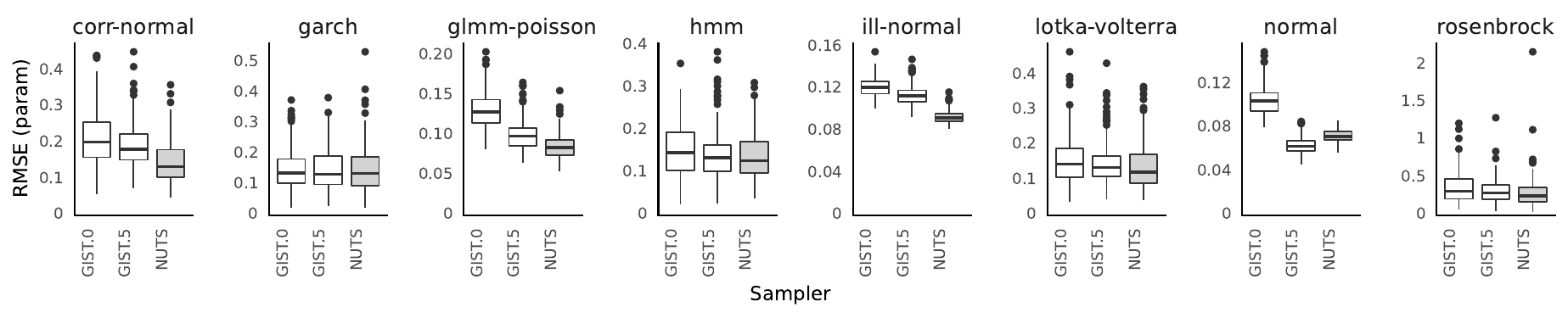}

    \includegraphics[width=0.95\textwidth]{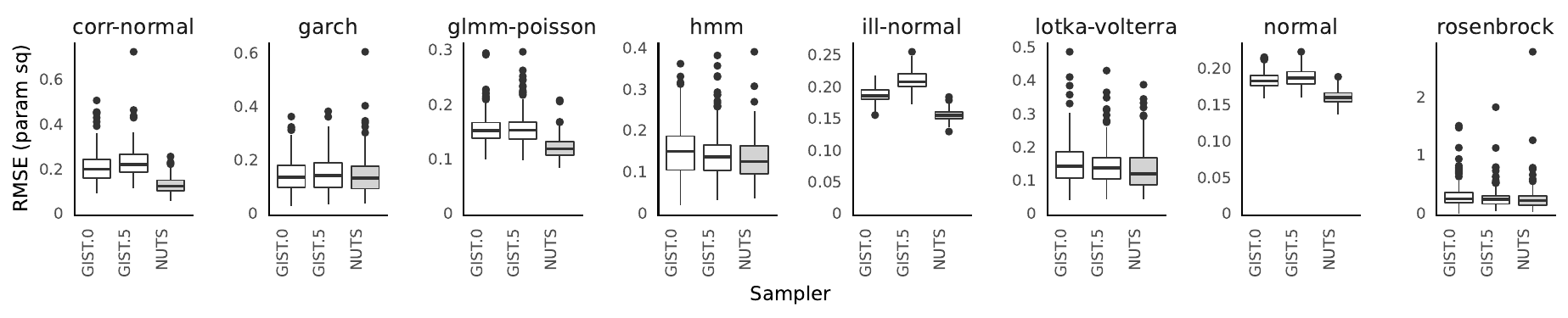}

    \includegraphics[width=0.95\textwidth]{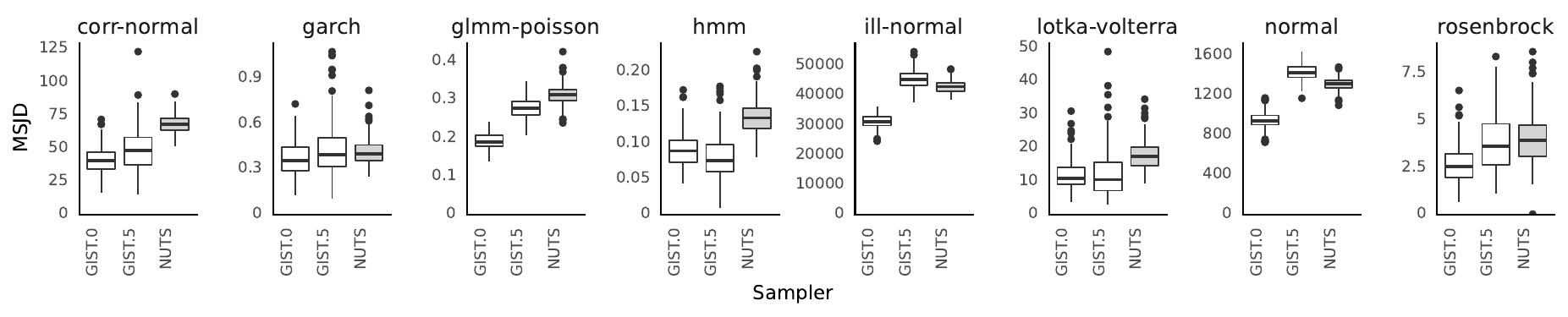}
    
    \includegraphics[width=0.95\textwidth]{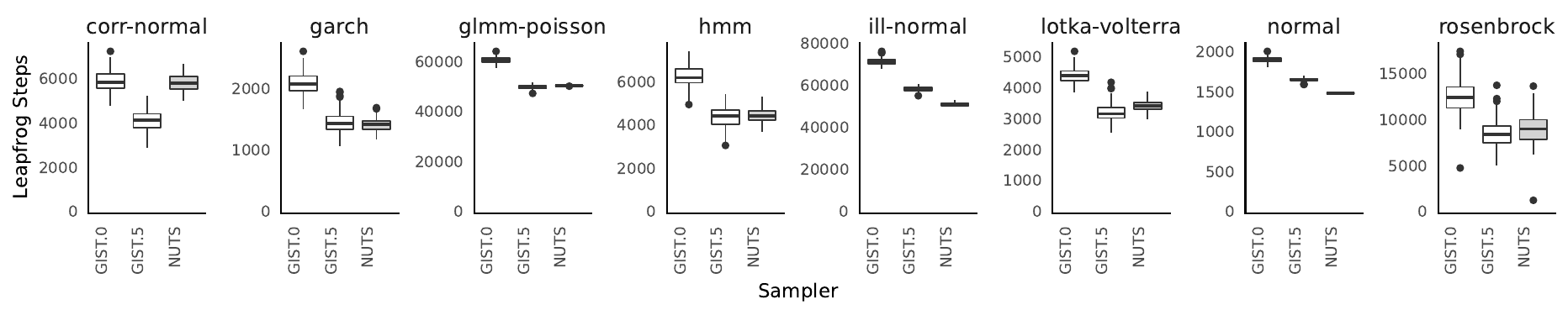}    
    \caption{\it {\bfseries Empirical error evaluations for GIST vs.~NUTS}.  Each column is a different model, with evaluations for RMSE for parameters, RMSE for parameters squared, MSJD, and number of leapfrog steps.  The Tukey-style bar-and-whisker plots report the median as a line, the 50\% central interval as a box, whiskers to min/max value within 1.5 times the central interval, and dots for individual outliers).  The GIST sampler is based on \eqref{eq:biased_selection_uniform_tuning_distribution} with  $\psi=0$ and $\psi = 0.5$, as indicated.}
    \label{fig:empirical-evals}
\end{figure}

To establish ground-truth estimates of the first and second moments, we ran all of our models for 25,000 warm-up and 50,000 sampling iterations of NUTS in a single chain using Stan's defaults and use the resulting draws to estimate parameters and parameters squared. To compare NUTS with GIST samplers based on \eqref{eq:biased_selection_uniform_tuning_distribution}, we ran 200 independent chains for 100 iterations from distinct initializations which are shared across all the samplers evaluated in the experiment. 

The RMSE for NUTS and GIST for parameters and parameters squared, along with MSJD and number of leapfrog steps are shown in Figure~\ref{fig:empirical-evals}.  GIST is evaluated with excluded proportions of $\psi=0$ and $\psi=0.5$.  Both NUTS and GIST use a unit mass matrix and the step size adapted by NUTS targeting a 90\% acceptance rate.

The top two plots in Figure~\ref{fig:empirical-evals} show RMSE for parameters (top) and parameters squared (bottom).  All plots have a $y$-axis starting at 0 so that relative performance may be assessed visually.  Many of the more complicated models are close in RMSE for parameters and parameters squared (i.e., first and second moments), with NUTS slightly outperforming GIST.  Nevertheless, the difference in mean performance between GIST and NUTS is typically much smaller than the variation of performance within GIST runs or within NUTS runs.

GIST with $\psi = 0.5$ excluded initial proportion requires fewer leapfrog steps than with a $\psi=0$ excluded initial proportion, as U-turns are hit sooner in the reverse direction.  With a $0.5$ path fraction, the number of steps between NUTS and GIST is comparable, with some models requiring fewer with GIST and some with NUTS.  As expected, error for parameter estimates (first row) is inversely related to mean square jump distance (third row).

In conclusion, NUTS only slightly outperforms the GIST sampler overall.  This is highly encouraging in that the GIST sampler is a relatively simple method, whereas NUTS uses Stan's current implementation, which has been implemented robustly and improved considerably over the first NUTS paper (and first implementation in Stan) \cite{betancourt2017conceptual,HoGe2014}.

\section{Other related prior work}

The Empirical HMC Sampler \cite{wu2018faster} learns an empirical distribution of the underlying path lengths to U-turns during a warmup phase and then fixes this empirical distribution, which is then used to sample path lengths at each step during sampling. Like randomized HMC in Section~\ref{ex:exact_rHMC}, the empirical HMC sampler can be cast as an instance of a GIST sampler where the conditional distribution of path lengths does not depend on the current position and/or momentum.

Both NUTS and the Apogee-to-Apogee Path Sampler, discussed in Section ~\ref{sec:pathlength}, can be formulated within the dynamic HMC framework introduced in \cite{durmus2023convergence}.  The GIST sampler generalizes this framework by incorporating a Metropolis step targeting the joint distribution in \eqref{eq:enlarged_target}.  A key advantage of the Metropolis step in GIST is that it relaxes the strict symmetry conditions typically required for reversibility of dynamic HMC, offering greater flexibility in tuning HMC's   parameters.

The autoMALA sampler introduced a self-tuning version of the Metropolis-adjusted Langevin (MALA) sampler \cite{kleppe2016adaptive,biron2024automala}.  It uses a forward and reverse non-deterministic scheme to choose adaptation parameters in a way that satisfies detailed balance. MALA is equivalent to one-step Hamiltonian Monte Carlo, but is simpler in not needing to evolve momentum.  The GIST sampler can be viewed as a probabilistic generalization of the autoMALA adaptation selection criteria.

For step size and mass matrix tuning, two approaches have been popular in practice. In the first, an adaptation phase is used to estimate algorithm tuning parameters. These tuning parameters are then fixed so that the resulting chain is Markovian. This is the strategy used by NUTS for step size and mass matrix adaptation for HMC \cite{HoGe2014}. In the second approach, the adaptation phase is never turned off, but the amount of adaptation is decreased so that asymptotically the results are valid.  This is the strategy used by delayed rejection Metropolis (DRAM) \cite{haario2006dram}.

Another approach to step-size adaptivity is to use delayed rejection HMC \cite{modi2023delayed}.  The delayed rejection algorithm \cite{mira2001metropolis,green2001delayed} is a generalization of Metropolis-Hastings to a sequence of proposal moves.  This proposal sequence can be tuned to start with larger scale moves and then scaled down for subsequent proposals \cite{haario2006dram}. In the same vein, the delayed rejection HMC sampler automatically tries smaller step sizes if proposals with larger step sizes are rejected, allowing it to sample from target densities with varying scale. Like other delayed rejection methods, it generates ``ghost points'' using reversed proposals, which are then used as part of the acceptance probability to ensure detailed balance \cite{green2001delayed}.

\section{Open-source code and reproducibility}

The results and plots can be reproduced from our Python code distribution, which is available from GitHub under a permissive open-source license.\footnote{The code is distributed under the MIT License at \url{https://github.com/bob-carpenter/adaptive-hmc}.}  The Stan models provide log densities and gradients through BridgeStan \cite{roualdes2023bridgestan}.  NUTS is run through CmdStanPy \cite{stan2024cmdstanpy}.

\section{Conclusion}

This paper presented a novel class of auxiliary variable methods for constructing locally adaptive HMC samplers called GIST -- signifying Gibbs self-tuning for HMC.  GIST not only generalizes the path-length adaptation strategy of the No-U-Turn Sampler (NUTS) but also unifies a broad class of locally adaptive HMC samplers, including NUTS, the Apogee-to-Apogee Path Sampler, multinomial HMC and randomized HMC. What makes this unification particularly compelling is that it reveals an underlying universality: when reversibility is preserved, the transition step of any locally adaptive HMC sampler naturally aligns with the transition step of a GIST sampler. Beyond this unification, the GIST framework unlocks: (i) simpler alternatives to NUTS for path-length tuning; but also (ii) local adaptation of NUTS’s other parameters \cite{BouRabeeCarpenterKleppeMarsden2024}. This makes GIST a powerful and versatile extension of NUTS, with broad implications for the future development of sampling algorithms for continuously differentiable densities.

\subsection*{Acknowledgements}

We would like to thank Andreas Eberle, Miika Kailas, Tore Kleppe, Sam Livingstone, Charles Margossian, Chirag Modi, Stefan Oberd\"{o}rster, Art Owen, Edward Roualdes, Gilad Turok, Aki Vehtari, and Matti Vihola
for feedback on Gibbs self tuning and an earlier draft of this paper.

N.~Bou-Rabee has been partially supported by NSF Grant No.~DMS-2111224.

\printbibliography

\appendix

\newpage

\section{Proofs}\label{app:proofs}

We begin by proving Lemma~\ref{lem:DeterministicMetropolization}.  While the result is surely well-known to experts \cite{tierney1998note},  we include a proof for completeness.

\begin{proof}[Proof of Lemma~\ref{lem:DeterministicMetropolization}]
Let $Z \sim \widehat{\mu}$ and $u \sim \operatorname{uniform}([0,1])$. Set $Z' \, = \, G(Z)$ and
\[
Z^* =\begin{cases}
Z' & \text{if  } u \le \dfrac{\widehat{\mu}( F(Z'))}{ \widehat{\mu}(F(Z))} \;,  \\
Z & \text{else} \;.
\end{cases}
\] The proof shows that 
\[
\prb(Z  \in A , Z' \in B) =  \prb(Z'  \in A , Z \in B) \;, 
\] 
for all measurable sets $A,B \subset \widehat{\mathbb{S}}$.

The  transition kernel $\widehat{\pi}$ in \eqref{eq:pi_det} satisfies \[
\widehat{\pi}(x,A) \ = \ \widehat{a}(x) \mathbb{1}_{A}(G(x)) + (1-\widehat{a}(x))\mathbb{1}_{A}(x) \;, \qquad \widehat{a}(x) = \min\!\left( 1, \frac{\widehat{\mu}(G(x))}{\widehat{\mu}(x)} \right) \;.
\] Hence,
\begin{align}
\prb( & Z  \in A , Z' \in B) = \int_A \widehat{\pi}(x,B) \widehat{\mu}(x) \lambda_{\widehat{\mathbb{S}}}(dx) \notag \\
&= \int_A \left[ \widehat{a}(x) \mathbb{1}_{B}(G(x)) + (1-\widehat{a}(x))\mathbb{1}_{B}(x) \right] \widehat{\mu}(x) \lambda_{\widehat{\mathbb{S}}}(dx) \label{eq:metrodef} \\
&= \int_{\widehat{\mathbb{S}}} \left[ \widehat{a}(x) \mathds{1}_B(G(x)) \mathds{1}_A(x) + (1 - \widehat{a}(x))\mathds{1}_A(x)\mathds{1}_B(x) \right] \widehat{\mu}(x) \lambda_{\widehat{\mathbb{S}}}(dx) \;. \label{eq:metropolisexpansion}
\end{align}
The second term in \eqref{eq:metropolisexpansion} is already symmetric in the sets $A,B$ --- so we turn our attention to the first term. For this term, we use the elementary identity $\mathsf{a} \min(1, \frac{\mathsf{b}}{\mathsf{a}}) = \mathsf{b} \min(1, \frac{\mathsf{a}}{\mathsf{b}})$ valid for all $\mathsf{a}, \mathsf{b} \ne 0$, as follows  \begin{align}
\int_{\widehat{\mathbb{S}}} & \widehat{a}(x) \mathds{1}_B(G(x)) \mathds{1}_A(x) \widehat{\mu}(x) \lambda_{\widehat{\mathbb{S}}}(dx)  \nonumber \\
%&= \int_\mathbb{S} \beta(x, F(x)) \mathds{1}_B(F(x)) \mathds{1}_A(x) \gamma(x) \lambda_{\widehat{\mathbb{S}}}(dx) \notag \\
&=\int_{\widehat{\mathbb{S}}} \mathds{1}_B(G(x)) \mathds{1}_A(x)  \widehat{\mu}(x) \min \left(1, \frac{\widehat{\mu}(G(x))}{ \widehat{\mu}(x)} \right) \lambda_{\widehat{\mathbb{S}}}(dx) \notag \\
&= \int_{\widehat{\mathbb{S}}} \mathds{1}_B(G(x)) \mathds{1}_A(x)  \widehat{\mu}(G(x)) \min \left(1, \frac{ \widehat{\mu}(x)}{\widehat{\mu}(G(x))} \right) \lambda_{\widehat{\mathbb{S}}}(dx)  \notag  \\
&=\int_{\widehat{\mathbb{S}}}  \widehat{a}(x) \mathds{1}_B(x) \mathds{1}_A(G(x))  \widehat{\mu}(x) \lambda_{\widehat{\mathbb{S}}}(dx) =\int_B \widehat{a}(x) \mathds{1}_A(G(x))  \widehat{\mu}(x) \lambda_{\widehat{\mathbb{S}}}(dx) \;.
\label{eq:metroaftercov}
\end{align}
where in the last step we used a change of variables under the $\lambda_{\widehat{\mathbb{S}}}$-preserving involution $G$.  Inserting (\ref{eq:metroaftercov}) into (\ref{eq:metropolisexpansion}), and comparing with (\ref{eq:metrodef}), we observe that
\[\prb(Z \in A , Z' \in B) = \prb(Z \in B , Z' \in A)\]
for all measurable sets $A,B \subset \widehat{\mathbb{S}}$, as required.
\end{proof}

With Lemma~\ref{lem:DeterministicMetropolization} in hand, we are now in position to prove reversibility of $\pi_{\mathrm{AV}}$ in \eqref{eq:piAV}.

\begin{proof}[Proof of Theorem~\ref{thm:auxvarmethod}]
Suppose  $  \theta \sim \mu$. After selecting  $v \sim p( \cdot \mid \theta)$, then $(\theta, v) \sim \widehat{\mu}$ where $\widehat{\mu}$ is given in \eqref{eq:joint}.  Let $u \sim \operatorname{uniform}([0,1])$. Set $(\theta', v') \, = \, G(\theta, v)$ and
\[
(\theta^*, v^*) =\begin{cases}
(\theta', v') & \text{if  } u \le \dfrac{\mu(\theta') p( v' \mid \theta')}{ \mu(\theta) p( v \mid \theta)} \;,  \\
(\theta, v) & \text{else} \;.
\end{cases}
\]

%Following this procedure, the dynamics of the $\theta$ marginal follow those given by the algorithm in Algorithm~\ref{algo:general-self-tuning-step}. 

Since $G$ is a measure-preserving involution, Lemma \ref{lem:DeterministicMetropolization} implies \begin{equation} \label{eq:conseq} \mathbb{P}((\theta, v) \in A, (\theta^*, v^*) \in B)  = \mathbb{P}( (\theta^*, v^*)\in A, (\theta, v) \in B)\end{equation}
for any measurable sets $A,B \subset \mathbb{S} \times \mathbb{V}$. In particular, for Borel sets $\Tilde{A}, \Tilde{B} \subset \mathbb{S}$,
\[ \mathbb{P}( \theta \in \Tilde{A}, \theta^* \in \Tilde{B}) = \mathbb{P}(\theta^* \in \Tilde{A}, \theta \in \Tilde{B})\]
by taking $A = \Tilde{A} \times \mathbb{V}$ and $B = \Tilde{B} \times \mathbb{V}$ in \eqref{eq:conseq}.  Hence, $\pi_{\mathrm{AV}}$ in \eqref{eq:piAV} is reversible with respect to  $\mu$.
\end{proof}

Theorem~\ref{thm:gist_reversibility} implies reversibility of the GIST samplers presented in Sections \ref{ex:exact_rHMC} and \ref{ex:exact_stHMC}, as well as the novel GIST sampler introduced in Section~\ref{sec:numericalHMC}.

\begin{proof}[Proof of Corollary \ref{cor:GIST-reversible}]
It suffices to verify that the map $G: \mathbb{R}^{2d} \times \mathbb{A} \to \mathbb{R}^{2d} \times \mathbb{A}$ defined by $G(\theta, \rho, \alpha) = (\mathcal{S} \circ F(\alpha)(\theta, \rho), \alpha)$ is a measure-preserving involution. As $\mathcal{S} \circ F(\alpha)$ is an involution, $G$ is automatically an involution on $\mathbb{R}^{2d} \times \mathbb{A}$. Additionally $\mathcal{S} \circ F(\alpha)$ preserves Lebesgue measure $m^{2d}$ on $\mathbb{R}^{2d}$ for every fixed $\alpha$.  Hence, by Fubini's theorem, $G$ preserves $(m^{2d} \otimes \eta)$ on $\mathbb{R}^{2d} \times \mathbb{A}$.  Thus, iterating the corresponding transition step in Algorithm~\ref{algo:general-self-tuning-step} produces a reversible Markov chain by Theorem~\ref{thm:gist_reversibility}.
\end{proof}

The next two corollaries of Theorem~\ref{thm:gist_reversibility} complete the proofs of reversibility of the GIST samplers in Section~\ref{sec:NUTS} and Section~\ref{sec:Apogee-to-Apogee}.

\begin{proof}[Proof of Corollary \ref{cor:NUTS-reversible}] 

In this case, the map $G$ is of the form $G(\theta, \rho, J, i) = (S \circ \Phi_h^i(\theta, \rho), -(J-i), i)$. This is clearly an involution. Moreover, for $\theta, \rho$ fixed the map $(J,i) \to (-(J-i), i)$ is a bijection and thus preserves the counting measure on $\mathcal{O} \times \mathbb{Z}$. By Fubini's theorem, $G$ then preserves $(m^{2d} \otimes \eta)$ on $\mathbb{R}^{2d} \times \mathcal{O} \times \mathbb{Z}$. Thus, by Theorem~\ref{thm:gist_reversibility}, we get reversibility of NUTS as presented in Section~\ref{sec:NUTS}. 
\end{proof}

\begin{proof}[Proof of Corollary \ref{cor:AAPS-reversible}] 
Here, $G(\theta, \rho, c, i)= \Big(\mathcal{S} \circ \Phi_h^i(\theta, \rho), c + S_{\#}((\theta, \rho), \mathcal{S} \circ \Phi_h^i(\theta, \rho)), i \Big)$. Since by definition $S_\#((\theta, \rho), (\theta', \rho')) = - S_\#((\theta', \rho'), (\theta, \rho))$, we observe by direct computation that $G$ is an involution. For fixed $(\theta, \rho, i)$, the map $c \mapsto c + S_{\#}((\theta, \rho), \mathcal{S} \circ \Phi_h^i(\theta, \rho))$ is a bijection and hence preserves the counting measure. Additionally, for $i$ fixed, $\theta, \rho \mapsto \mathcal{S} \circ \Phi_h^i(\theta, \rho)$ preserves Lebesgue measure on $\mathbb{R}^{2d}$. Applying Fubini's theorem then implies that the map $G$ is measure preserving. Hence, by Theorem~\ref{thm:gist_reversibility}, the Apogee-to-Apogee Path Sampler presented in Section~\ref{sec:Apogee-to-Apogee} is reversible. \end{proof}

\newpage

\section{Test models}

\label{sec:test_models}

In this section, we briefly describe the models used for evaluations.  Greek letters are used for parameters, Roman letters for constants, predictors and modeled data, and italics for indexes.  Where not otherwise specified, parameters have weakly informative priors concentrated on their rough scale.

\begin{description}
\small
\item[Standard normal]
A 500-dimensional standard normal, with $\alpha \sim \textrm{normal}(0, \textrm{I}_{500 \times 500})$.

\item[Correlated normal]
A 250-dimensional correlated normal $\alpha \sim \textrm{normal}(0, \textrm{S}),$ where the covariance is that of a unit scale first-order random walk, $\textrm{S}_{i, j} = \textrm{r}^{|i - j|}$, with correlation $\textrm{r} = 0.9$.

\item[Ill-conditioned normal]
A 250-dimensional ill-conditioned normal, with $\alpha \sim \textrm{normal}(0, \textrm{diag-matrix}(\sigma)),$ where $\sigma = \left[ \frac{1}{250} \ \ \frac{2}{250} \cdots \frac{250}{250}\right]^\top.$

\item[Poisson generalized linear mixed model]
The expected count for observation $i$ is
$
\lambda_i =  \exp( \alpha + \beta_1 \cdot \textrm{t}_i + \beta_2 \cdot \textrm{t}_i^2 + \beta_3 \cdot \textrm{t}_i^3 + \varepsilon_i),
$
with a hierarchical prior
$
    \varepsilon_i \sim \textrm{normal}(0, \sigma),
$
and a likelihood
$
    y_i \sim \text{Poisson}(\lambda_i),
$
for $i < 40.$

An autoregressive time series with first element $y_1 \sim \textrm{normal}(\mu + \phi \cdot \mu, \sigma),$ and subsequent elements $y_t \sim \textrm{normal}(\mu + \phi \cdot y_{t-1} + \theta \cdot \epsilon_{t-1}, \sigma)$ for $t > 1.$

\item[Generalized autoregressive conditional heteroskedasticity (GARCH)] 
An autoregressive time series model with stochastic volatility, with $y_t \sim \textrm{normal}(\mu, \sigma_t)$, where $\sigma_1 = s$ is given as data and \[ \sigma_t = \sqrt{\alpha_0 + \alpha_1 \cdot (y_{t-1} - \mu)^2 + \beta_1 \cdot \sigma_{t-1}^2}.
\]

\item[Hidden Markov model]
A hidden Markov model (HMM) with normal observations.  The data generating process is Markovian in hidden state, $z_t \sim \textrm{categorical}(\phi_{z_{t-1}}),$ and then normal in observation, $y_t \sim \textrm{normal}(\mu_{z_t}, \sigma_{z_t}).$  In the implementation, the forward algorithm is used to marginalize the $z_t$ to calculate the likelihood.  The vector $\mu$ is constrained to ascending order for identifiability.

\item[Lotka-Volterra population dynamics]  A lognormal model of population time series for prey ($y_{t,1}$) and predator ($y_{t,2}$).  The population dynamics is modeled by a system of ordinary differential equations, $\frac{\textrm{d}}{\textrm{d}t} u(t) = (\alpha - \beta \cdot v(t)) \cdot u(t)$ and $\frac{\textrm{d}}{\textrm{d}{t}} v(t) = (-\gamma + \delta \cdot u(t)) \cdot v(t)$ with unknown starting point $(u(0), v(0))$ and discrete observations modeled by $y_{t, 1} \sim \textrm{lognormal}(\log u(t), \sigma_1)$ and $y_{t, 2} \sim \textrm{lognormal}(\log v(t), \sigma_2),$  where $u(t)$ and $v(t)$ are solutions to the ODE.

$\frac{\textrm{d}}{\textrm{d}t} C
= \frac{\nu}{V} \cdot \frac{C(t)}{\mu + C(t)}
- \exp(-\delta \cdot t) \cdot D \cdot \frac{\delta}{V},$ with lognormal likelihood for discrete observations
$y_t \sim \textrm{lognormmal}(\log C(t), \sigma),$ with all parameters constrained to be positive.

\item[Rosenbrock distribution]  A two-dimensional ``banana'' distribution with $v \sim \textrm{normal}(1, 1)$ and $\theta \sim \textrm{normal}(v^2, 0.1).$

\end{description}
\end{document}